\documentclass[a4paper,11pt]{article}

\usepackage{jheppub}
\usepackage{latexsym,amsmath,amssymb,mathrsfs,hyperref}
\usepackage{slashed}
\usepackage{tensor}
\usepackage{graphicx}
\usepackage{tikz}
\usepackage[T1]{fontenc}
\usepackage{color}
\newcommand{\beq}{\begin{equation}}
\newcommand{\eeq}{\end{equation}}
\newcommand{\IP}{{\mathscr I}^+}
\newcommand{\IM}{{\mathscr I}^-}

\hypersetup{
	colorlinks=true,
	linkcolor=blue,
	filecolor=magenta,      
	urlcolor=blue,
	citecolor=blue
}
\numberwithin{equation}{section}

\makeatletter
\gdef\@fpheader{}
\makeatother

\preprint{NORDITA 2020-104}
	
	\title{\centering \huge Super-Hawking Radiation}
	
		\author[a,b]{Ricardo Z. Ferreira}
	\author[b,c]{and Carlo Heissenberg}

	\affiliation[a]{Institut de Fisica d’Altes Energies (IFAE) and  The Barcelona Institute of Science and Technology (BIST), Campus UAB, 08193 Bellaterra, Barcelona.}
		\affiliation[b]{Nordita, KTH Royal Institute of Technology and Stockholm University, \\ Roslagstullsbacken 23, SE-106 91 Stockholm, Sweden}
		\affiliation[c]{Department of Physics and Astronomy, Uppsala University, \\ Box 516, SE-751 20 Uppsala, Sweden}
		
		\emailAdd{rzambujal@ifae.es}\emailAdd{carlo.heissenberg@su.se}

\abstract{
	We discuss modifications to the Hawking spectrum that arise when the asymptotic states are supertranslated or superrotated. 
	For supertranslations we find nontrivial off-diagonal phases in the two-point correlator 
	although the emission spectrum is eventually left unchanged, as previously pointed out in the literature.
	In contrast, superrotations give rise to modifications which manifest themselves in the emission spectrum and depend nontrivially on the associated conformal factor at future null infinity.
	We study Lorentz boosts and a class of superrotations whose conformal factors do not depend on the azimuthal angle on the celestial sphere and whose singularities at the north and south poles have been associated to the presence of a cosmic string. 
	In spite of such singularities, superrotations still lead to finite spectral emission rates of particles and energy which display a distinctive power-law behavior at high frequencies for each angular momentum state. 
	The integrated particle emission rate and emitted power, on the contrary, while finite for boosts, do exhibit ultraviolet divergences for superrotations, between logarithmic and quadratic. 
	Such divergences can be ascribed to modes with support along the cosmic string. In the logarithimic case, corresponding to a superrotation which covers the sphere twice, the total power emitted still presents the Stefan-Boltzmann form but with an effective area which diverges logarithmically in the ultraviolet.}

\begin{document}
		\maketitle		
	
	\section{Introduction}
	
	The asymptotic symmetry group of asymptotically flat spacetimes, the BMS group, made its first appearance long ago and was named after its authors Bondi, Metzner, van der Burg and Sachs \cite{Bondi:1962px, Sachs:1962wk, Sachs:1962zza} (see e.g. \cite{Geroch:1977jn,Wald:1984rg,Ashtekar:1987tt} for  introductory presentations). 
	Perhaps surprisingly at the time, this group was found to contain, together with the Lorentz group, an infinite-dimensional subgroup of enhanced translations known as supertranslations. 
	More recently, motivated in particular by  outstanding progress in the context of two-dimensional conformal field theories, the structure of infinitesimal BMS transformations was revisited by Barnich and Troessaert, who proposed a natural extension of the Lorentz algebra to two copies of the Virasoro algebra \cite{Barnich:2009se,Barnich:2010eb,Barnich:2011ct}, identifying an infinite-dimensional family of superrotations. This extension is based on the fact that boosts and rotations act on asymptotically flat spacetimes via the identification between the Lorentz group and the group of globally well-defined conformal transformations on the celestial two-sphere $SO(1,3)\simeq SL(2,\mathbb C)\simeq \mathrm{Conf}(S^2)$. In contrast with standard boosts and rotations, superrotations typically feature singularities on the celestial sphere such as poles and branch cuts. An even further extension of superrotations to arbitrary transformations on the sphere has been put forward by Campiglia and Laddha \cite{Campiglia:2015yka}, based on suitably relaxing the BMS boundary conditions, although we here focus for definiteness on the Barnich--Troessaert superrotations.
	 
	Asymptotic symmetries have also experienced a revival due to the unveiling of unexpected connections with observable effects. On the one hand, soft theorems have been recast as Ward identities for asymptotic symmetry transformations, not only for scattering amplitudes on (asymptotically) flat spacetime \cite{Yennie:1961ad,Weinberg_64,Weinberg_65,Strominger2014,He2015,soft-subleading,Broedel:2014fsa,Bern:2014vva,He:2014cra,Campiglia:2014yka,Campiglia:2015qka} but also in the case of correlation functions  on (anti-) de Sitter background \cite{Balasubramanian:1999re,deHaro:2000vlm,Anninos:2010zf,Creminelli:2012ed,Hinterbichler:2013dpa,Ferreira:2016hee,Ferreira:2017ogo,Compere:2019bua,Compere:2020lrt}.
	On the other hand, directly observable counterparts of asymptotic symmetries have been identified in the so-called memory effects, permanent footprints that radiation can leave behind on a test apparatus \cite{Zeldovich:1974gvh,Christodoulou:1991cr,Bieri:2013hqa,Strominger2016}. These observations provided a deeper understanding of the nature of such symmetries, exposing concretely why they should not be regarded as trivial redundancies, but actually as transformations between inequivalent field and matter configurations. 
	This picture of asymptotic symmetries has also been extended to higher-dimensional setups  \cite{Kapec:2014zla,Kapec:2015vwa,Pate:2017fgt,Aggarwal:2018ilg,Henneaux:2019yqq,Freidel:2019ohg,Campoleoni:2019ptc} and higher-spin gauge theories \cite{Campoleoni:2017mbt,Campoleoni:2017qot,Campoleoni:2018uib}.
	In addition, a nontrivial interplay between asymptotic symmetries and dualities has been pointed out \cite{Strominger-dual,Shahin-dual,Campiglia-dual,twoform-dual,GodazgarDual1,GodazgarDual2,Henneaux:2018mgn,PorratiDual,Henneaux:2020nxi} thus providing an additional piece of theoretical appeal to the subject. Asymptotic symmetries, soft theorems and memory effects thus comprise the three corners of a recurring structure known as an infrared triangle, many instances of which  appear in different contexts in gauge theories and gravity (see \cite{Strominger:2017zoo} for a review).
	
The interest in asymptotic symmetries is further motivated by their potential ties to the black-hole information paradox. Soon after discovering that the formation of a black hole gives rise to the creation of a thermal spectrum of emitted particles as seen by a faraway observer \cite{Hawking1975}, Hawking himself realized that this process, especially after the subsequent complete evaporation of the black hole, gives rise to a seeming contradiction with the principle of unitary evolution in quantum mechanics \cite{Hawking:1976ra}. 
The realization that spacetimes related by an asymptotic symmetry should not be identified but rather regarded as physically inequivalent, despite being linked by a diffeomorphism albeit a ``large'' one, has stimulated a critical revision of the basic assumptions of the no-hair theorem and has lead to the identification of additional conserved quantities that should be employed to specify a black-hole configuration: the supertranslation and superrotation charges, often referred to as ``soft hair'' decorating the black-hole horizon \cite{Hawking2016,Hawking2017,Haco:2018ske,Haco:2019ggi}.
The actual physical relevance of such charges and the extent to which they indeed give rise to appropriate labels for black-hole states has also been critically investigated in  \cite{PorratiMir,PorratiWig,Gomez:2017ioy, Chu:2018tzu} (see also \cite{Lin:2020gva} for a discussion of nonlinearities in lower-dimensional theories).

If these additional charges are indeed to bear relevance to the problem of information loss, one expects that the Hawking spectrum itself should be sensitive to the action of asymptotic symmetries. 	
The effect of supertranslations on Hawking radiation has been studied, in particular, in two recent papers \cite{Javadinezhad2019,Compere2019}, where it was shown that an asymptotic supertranslation indeed gives rise to an angular mixing in the Bogolyubov coefficients connecting early- and late-time observers, while leaving the Hawking spectrum unchanged, as already anticipated by Hawking in his seminal paper \cite{Hawking1975}. 
The case of superrotations is more intriguing because, except for standard boosts and rotations, such transformations can introduce point- or string-like singularities on the celestial sphere at null infinity. As a result, they preserve asymptotic flatness only locally, while they break it at the global level. Point singularities due to superrotations have been related to the appearance of cosmic strings \cite{Strominger2017} or to an effective deformation of the celestial sphere to an elongated object, a ``cosmic football'' \cite{Adjei:2019tuj}.
	
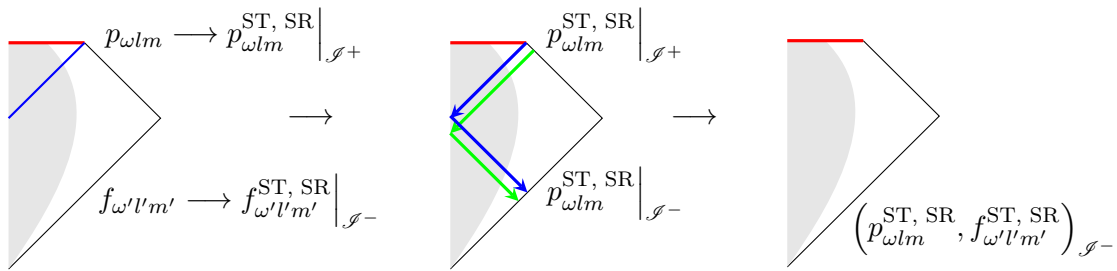
\begin{figure}[h!]
	\centering
	\begin{equation*}
	\begin{gathered}
	\begin{tikzpicture}
	\path [draw] (0,0)--(2,2)--(1,3);
\path [fill=black!10!white] (0,0) to[out=55,in=-45] (.5,3)--(0,3)--(0,0);	
\draw[red, very thick] (0,3)--(1,3);  
\draw[,blue, thick] (1,3)--(.0,2);
				\node at (1.1,3.1)[right]{$ \left. p_{\omega l m}  \longrightarrow   p^\text{ST, SR}_{\omega l m} \right|_{\IP}$};
								\node at (2.0,2.0)[right]{$ \qquad \qquad  \longrightarrow$};
\node at (1.0,0.9)[right]{$\left. f_{\omega' l' m'} \longrightarrow f^\text{ST, SR}_{\omega' l' m'} \right|_{\IM} $};
	\end{tikzpicture}
\end{gathered}
	 \quad \quad 
	\begin{gathered}
	\begin{tikzpicture}
	\path [draw] (0,0)--(2,2)--(1,3);
\path [fill=black!10!white] (0,0) to[out=55,in=-45] (.5,3)--(0,3)--(0,0);	
\draw[red, very thick] (0,3)--(1,3);
		\draw[-stealth,blue,very thick] (1,3)--(.0,2);
		\draw[-stealth,green,very thick] (1.1,2.9)--(.0,1.8);
				\draw[-stealth,green,very thick] (.0,1.8)--(0.9,0.9);
		\draw[-stealth,blue,very thick] (0,2.03)--(1.015,1.015);
				\node at (2.0,2.0)[right]{$ \qquad   \longrightarrow$};
		\node at (1.1,3.1)[right]{$ \left. p^\text{ST, SR}_{\omega l m} \right|_{\IP}$};
		\node at (1.1,1.0)[right]{$\left. p^\text{ST, SR}_{\omega l m}\right|_{\IM}$};
	\end{tikzpicture}
\end{gathered}
	\quad \quad 
		\begin{gathered}
			\vspace{-0.4cm}
		\begin{tikzpicture}
			\path [draw] (0,0)--(2,2)--(1,3);
			\path [fill=black!10!white] (0,0) to[out=55,in=-45] (.5,3)--(0,3)--(0,0);	
			\draw[red, very thick] (0,3)--(1,3);
		\node at (1.1,3.0)[right]{};
			\node at (0.7,0.5)[right]{$\left(p^\text{ST, SR}_{\omega l m} ,f^\text{ST, SR}_{\omega' l' m'}\right)_{\IM}$};
		\end{tikzpicture}
	\end{gathered}
\end{equation*}
	\caption{\label{Diagramatic picture of the computation} Diagrammatic picture of the computation performed in this work. From left to right: We first transform the free-particle states at future and past null infinity by an asymptotic symmetry (supertranslation or superrotation). We then propagate the transformed state at future infinity back to past infinity. Finally, we find the Hawking spectrum by projecting such state into the transformed state at past null infinity. }
	\label{}
\end{figure}

Keeping in mind these possible shortcomings associated to superrotations, in this work we investigate whether the Hawking spectrum is corrected if one allows the asymptotic states at past ($\IM$) and future  ($\IP$) null infinity to be supertranslated or superrotated. More concretely, we start by considering a massless, minimally coupled scalar field on the background of a spherically symmetric gravitational collapse, i.e. Hawking's original setup. The associated free single-particle states $p_{\omega l m}$ ($f_{\omega l m}$) at $\IP$ ($\IM$) can be thus labeled by their energy and angular momentum, or equivalently by their asymptotic four-momentum,  $p_{\omega \hat q}$ ($f_{\omega' \hat q'}$) at $\IP$ ($\IM$). We then perform a finite BMS transformation (i.e. a finite supertranslation or a boost/superrotation) of these asymptotic states and compute the associated Bogolyubov coefficients while leaving the bulk geometry and dynamics unchanged, in particular without spoiling their spherical symmetry. 
These coefficients provide access to the two-point functions $\langle 0_-|\boldsymbol b_{\omega  l  m}^\dagger \boldsymbol{b}_{\omega'  l' m'} |0_-\rangle $ or $\langle 0_-|\boldsymbol b_{\omega  \hat q}^\dagger \boldsymbol{b}_{\omega'  \hat q'} |0_-\rangle $ involving the ladder operators $\boldsymbol b$, $\boldsymbol b^\dagger$ defined by the asymptotic observer at $\mathscr I^+$ and the vacuum state $|0_-\rangle$ defined by the asymptotic observer at $\mathscr I^-$. The diagonal entries of the two-point functions then characterize the spectral emission rates, namely the number of particles $dN_{lm}(\omega)$ emitted to $\mathscr I^+$ per unit time and frequency for each $l$, $m$, or the analogous rate $dN(\omega,\hat q)$ per unit time, frequency and solid angle, and the emitted power spectrum $dP_{lm}(\omega)=\omega \, dN_{lm}(\omega)$ or $dP(\omega,\hat q)=\omega\, dN(\omega, \hat q)$. A schematic illustration of the calculation is given in Figure~\ref{Diagramatic picture of the computation}.

Following this strategy we find that supertranslations do in general modify both the Bogolyubov coefficients and the two-point function by inducing nontrivial off-diagonal phases which depend on the supertranslation acting at $\mathscr I^+$. However, they eventually leave the spectral emission rate $dN(\omega, \hat q)$ unaltered.
In the case of superrotations, we show that the Bogolyubov coefficients, the two-point function and the emission spectra all depend nontrivially on the associated conformal transformations and conformal factors at $\mathscr I^+$. This aligns with the expectation that, while ordinary rotations clearly cannot alter Hawking radiation for a spherically symmetric setup, already Lorentz boosts ought to affect the spectrum via Doppler effect. 
Both for supertranslations and for superrotations, the two-point function is only sensitive to the asymptotic symmetry at future null infinity, while the dependence on the corresponding transformation at past null infinity drops out.

For the Barnich--Troessaert superrotations we find strong corrections with respect to thermality that arise from the region on the celestial sphere where the associated conformal factor $R_+(\hat q)$ 
tends to zero, which therefore comprise an intrinsic signature of superrotations. In particular, the spectral rate $dN(\omega,\hat q)$ diverges for directions $\hat q$ aligned with the zeros of $R_+(\hat q)$.
Let us stress that, despite this singular behavior, the spectral emission rate $dN_{lm}(\omega)$, 
is not dominated by the point-like singularities, but rather receives its leading contributions from a smooth region around them, which leads to a well-behaved and finite result. Similar features are also exhibited by the power spectrum.
On the contrary, the total, integrated particle emission rate and emitted power, while of course finite for boosts with a given velocity $v<1$, generically exhibit divergences for superrotations.

In order to make the above features more explicit, in some calculations we will focus for definiteness on superrotations described by the conformal transformations $(z,\bar z)\mapsto c\,\left(z^n,\bar z^n\right)$ in stereographic coordinates $(z,\bar z)$, where $n$ is a positive integer and $c>0$ can be regarded as a rapidity. Boosts along the third spatial direction correspond to $n=1$, while larger values of $n$ correspond to {bona fide} superrotations.
In the pure-absorption approximation, we find that the expansion of the particle emission spectrum $dN_{lm}(\omega)$ for large frequencies, ${2\pi\omega}/{\kappa}\gg 1,c$ with $\kappa$ the black hole surface gravity, still exhibits the familiar exponential decay in $\omega$. On the contrary, whenever $n=2,3,4,\ldots$ the superrotated spectrum of emitted particles, while still finite, is characterized by a power-law decay at large frequencies. This behavior then leads to important corrections to the spectral emission rate, the emitted power spectrum and to the corresponding integrated quantities. We obtain that superrotations with $n=2$ lead to a total emitted power which satisfies the Stefan-Boltzmann law for a radiating body but with a rescaled effective area which diverges logarithmically in the ultraviolet. On the other hand, for $n>2$ the emitted power shows subquadratic divergences. Amusingly enough, the total number of particles emitted per unit time is instead finite for $n=2$ while exhibiting a logarithmic divergence for $n=3$ and sublinear divergences for larger $n$.

The paper is organized as follows. In Section~\ref{sec: Hawking spectrum} we review Hawking's original derivation of particle creation by black holes in order to establish the notation and highlight a few points that play an important role in the subsequent parts of the paper. After a brief recap of finite BMS transformations in Section~\ref{sec:BMSreview}, we then derive the general properties of supertranslated and boosted/superrotated spectra in Section~\ref{sec: supertranslations} and Section~\ref{sec: superrotations}.
We specialize to the class of boosts and superrotations outlined above in order to present some explicit results concerning the high-energy properties of boosted spectra, $2\pi\omega/\kappa\gg 1,c$ and $n=1$, in Subsection~\ref{ssec: HELboosts} and of superrotated spectra, $2\pi\omega/\kappa\gg1,c$ and $n=2,3,\ldots$, in Subsection~\ref{ssec: HELsuperrotations}.  In Subsection~\ref{sec:URBoost} we study the case of an ultrarelativistic boost $c\gg2\pi\omega/\kappa\gg1$ where some features, analogous in spirit to the ones highlighted for superrotations, also appear. A discussion of the total emitted power and total particle emission rate is then presented in Section \ref{sec: power spectrum}, first for generic boosts/superrotations and then for transformations of the form $(z,\bar z)\mapsto c\,\left(z^n,\bar z^n\right)$.

The paper also contains a few appendices. Appendix~\ref{app:basics} collects standard material concerning quantization on curved backgrounds and solutions of the wave equation on flat and Schwarzschild backgrounds. 
In Appendix~\ref{app: spherical harmonics} we summarize our conventions for spherical harmonics and related functions, while also recalling a few useful formulas.
Appendix~\ref{app: Geometry of spherical collapse} reviews the propagation of null rays on the background of a gravitational collapse. In Appendix~\ref{app:conformal} the properties of some relevant conformal transformations are discussed. Appendix~\ref{app:R=0} is devoted to the analysis of certain contributions arising from points where the conformal factor vanishes. Finally Appendix~\ref{app: Transmission coefficients} discusses the frequency behavior of the  transmission coefficients, and their associated density of states, for massless minimally coupled scalar fields.

\section{Hawking Spectrum \label{sec: Hawking spectrum}}

In order to setup the notation and highlight some features that will be relevant in the ensuing discussion, let us review Hawking's original derivation \cite{Hawking1975} of the black-hole emission spectrum in some detail. 

\begin{figure}[h!]
	\centering
	\begin{tikzpicture}
		\path [draw] (0,0)--(4,4)--(2,6);
		\path [fill=black!10!white] (0,0) to[out=55,in=-45] (.5,6)--(0,6)--(0,0);	
		\draw[red, very thick] (0,6)--(2,6);	
		\path [draw,green!60!black] (1.8,1.8)--(0,3.6)--(2.2,5.8);
		\path [draw,green!60!black] (1.9,1.9)--(0,3.8)--(2.1,5.9);
		\path [draw,green!60!black] (1.95,1.95)--(0,3.9)--(2.05,5.95);
		\draw[green!60!black,very thick] (2.015,2.015)--(0,4.03);
		\draw[blue,very thick] (2,6)--(.0,4);
		\node at (2.8,5.3)[right]{$\mathscr I^+$};
		\node at (2.8,2.8)[right]{$\mathscr I^-$};
		\node at (4,4)[right]{$i^0$};
	\end{tikzpicture}
	\caption{Penrose diagram of a gravitational collapse and subsequent formation of a black hole, whose event horizon  is depicted in blue. The grey area represents the interior of the collapsing matter, while the red line corresponds to the black-hole singularity. The green lines are null hypersurfaces, with the thick green line corresponding to $v=v_0$.}
	\label{fig:PenroseBHC}
\end{figure}
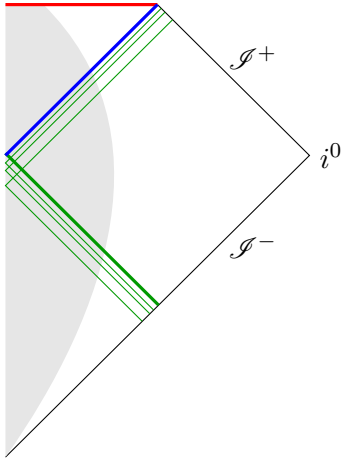

Hawking's goal was to compare the vacuum states at past and future null infinity
for a massless scalar field propagating in the spacetime depicted in Figure~\ref{fig:PenroseBHC}: a spherically symmetric gravitational collapse inducing the formation of a black hole.
Due to the impossibility of unambiguously defining positive and negative frequencies in curved space, these two vacua can look quite different. 

To get started, let us therefore consider the quantization of the free massless scalar field $\boldsymbol \phi$ following the standard strategy, a brief account of which is given in Appendix \ref{app:quantization-scalar}. We can first write
\beq
\boldsymbol{\phi}=\sum_{lm}  \int_0^\infty d\omega  \, \left(f_{\omega l m} \boldsymbol{a}_{\omega l m}+\bar{f}_{\omega l m} \boldsymbol{a}_{\omega l m}^{\dagger}  \right)\,,
\eeq 
where $ \boldsymbol{a}^\dagger_{\omega l m}$, $\boldsymbol a_{\omega l m}$ create and annihilate the single-particle states $f_{\omega l m}$, labeled by their frequency and angular momentum, as defined by an observer at past null infinity $\IM$.
The vacuum state for such an observer $|0_-\rangle$ is then characterized by 
\begin{equation}\label{}
	\boldsymbol  a_{\omega l m} |0_-\rangle=0\,.
\end{equation}  
Alternatively, the field can be expanded as
\beq
\boldsymbol{\phi}=\sum_{lm}   \int_0^\infty d\omega  \, \left(p_{\omega l m} \boldsymbol{b}_{\omega l m}+\bar{p}_{\omega l m} \boldsymbol{b}_{\omega l m}^{\dagger} \right)
\eeq
in terms of single-particle states $p_{\omega l m}$ as defined at future null infinity, $\IP$, where the vacuum is thus defined by
\begin{equation}\label{}
	\boldsymbol b_{\omega l m}|0_+\rangle=0\,.
\end{equation}
The associated Bogolyubov coefficients are then given by
\begin{align}\label{Bogcont}
	\alpha_{\omega\omega'll'mm'} = (p_{\omega l m}, f_{\omega' l' m'})\,,\qquad
	\beta_{\omega\omega'll'mm'} = - (p_{\omega l m}, \bar f_{\omega' l' m'})
\end{align}
in terms of the invariant Klein--Gordon scalar product.
Therefore, in general, the vacuum state $|0_-\rangle$ will not look empty to an observer at $\IP$, but rather it will appear populated by a spectrum of particles as quantified by the two-point function
\beq\label{corrn}
\langle 0_-| \boldsymbol b_{\omega l m}^\dagger \boldsymbol b_{\omega'l'm'} |0_-\rangle =\sum_{l'',m''} \int_0^{\infty} d\omega''\, \beta_{\omega \omega''l l'' m m''}  \bar\beta_{\omega' \omega'' l' l'' m' m''}\, .
\eeq 
In the following, angular brackets $\langle\,\cdots \rangle$ will stand as a shorthand for $\langle 0_-|\cdots|0_-\rangle$.

The mode functions $p_{\omega l m}$ and $f_{\omega l m}$ for the spherically symmetric setup under consideration can be written as follows,
\beq \label{solutions zero charge}
p_{\omega l m}(u,r,\hat x) =\frac{P_{\omega l}(r)}{\sqrt{2\pi\omega}\,r}\, e^{i\omega u} \,Y_{lm}(\hat x)\,,\qquad
f_{\omega l m}(v,r,\hat x) =\frac{F_{\omega l}(r)}{\sqrt{2\pi\omega}\,r}\, e^{i\omega v} \, Y_{lm}(\hat x)\, ,
\eeq
where $u$ and $v$ are the retarded and advanced time coordinates, $P_{\omega l}(r)$ and $F_{\omega l}(r)$ are suitable radial functions (see Appendices~\ref{app:wave-flat} and \ref{app:wave-Schwarzschild} for more details) and $Y_{lm}(\hat x)$ are the standard spherical harmonics (see Appendix~\ref{app: spherical harmonics} for a summary of conventions and a few useful formulas).
They indeed identify the proper free-particle states as seen by the observers at $\mathscr I^+$ and $\mathscr I^-$, since their asymptotic behavior there is
\begin{align}
	\label{limp+}
	p_{\omega l m}(u,r,\hat x)\,&\sim\, \frac{1}{\sqrt{2\pi\omega}\, r}\, e^{i\omega u}\, Y_{lm}(\hat x)\qquad (\text{at }\mathscr I^+)\,,\\
	\label{limf-}
	f_{\omega l m}(v,r, \hat x) \,&\sim\, \frac{1}{\sqrt{2\pi\omega}\, r}\, e^{i\omega v}\, Y_{lm}(\hat x) \qquad (\text{at }\mathscr I^-)\,.
\end{align}
One can check that they are also orthonormal with respect to the Klein--Gordon scalar product. Performing the integrals on $\mathscr I^-$ as in \eqref{SP} one has, for instance,
\begin{equation}\label{}
	(f_{\omega lm}, f_{\omega'l'm'}) =
	\frac{\omega+\omega'}{2\sqrt{\omega \omega'}} \int \frac{dv}{2\pi}\,  e^{iv(\omega-\omega')}\oint d\Omega(\hat x) Y_{lm}(\hat x)Y^\ast_{l'm'}(\hat x)= \delta(\omega-\omega')\delta_{ll'} \delta_{mm'}\,,
\end{equation}
while
\begin{equation}
	(f_\omega, \bar f_{\omega'})=0\,,\qquad
	(\bar f_\omega, \bar f_{\omega'})=-\delta(\omega-\omega')\delta_{ll'} \delta_{mm'}
\end{equation} 
and similar relations can be conveniently obtained for $p_{\omega l m}$ evaluating the integrals at $\mathscr I^+$.

To identify the appropriate asymptotic limit of $p_{\omega l  m}(u,r,\hat x)$ at \emph{past} null infinity, one can instead resort to geometrical optics. As reviewed in Appendix \ref{app: Geometry of spherical collapse}, a light ray emitted from $\mathscr I^-$ at an advanced time $v$  reaches $\mathscr I^+$ at a retarded time $u$ given by
\begin{equation}\label{u(v)}
	u(v)=-\frac{1}{\kappa}\log\frac{v_0-v}{C}\,.
\end{equation}  
This relation holds for $v$ close to $v_0$, the last retarded time at which a light ray emitted from $\mathscr I^-$ propagates through the collapsing matter and escapes to $\mathscr I^+$ before the horizon formation (see Figure~\ref{fig:PenroseBHC}). In eq.~\eqref{u(v)}, $\kappa=(2r_s)^{-1}$ is the black hole surface gravity and  $C$ is a constant which depends on the details of the collapse.
In this approximation,
\begin{equation}\label{limp-}
	p_{\omega l m}(u,r,\hat x)\,\sim\, \frac{t_{\omega l}}{\sqrt{2\pi\omega}\,r}\,e^{i\omega u(v)}
	\Theta(v_0-v)Y_{lm}(\hat x)\qquad (\text{at }\mathscr I^-)\,,
\end{equation}
where $\Theta$ denotes the Heaviside step function.
Here $t_{\omega l}\equiv P_{\omega l}(r_s)$, with $r_s$ the Schwarzschild radius, plays the role of a transmission amplitude. It quantifies the fact that the above estimate only captures the portion of the mode function that actually travels through the star.
To confirm this interpretation, one may quantify the amount by which $p_\omega$ fails to be properly normalized when evaluated at $\mathscr I^-$ according to \eqref{limp-}. Following steps very similar to those which we are going to detail below, one finds
\begin{equation}
	(p_{\omega l m}, p_{\omega'l'm'})_{\mathscr I^-} = |t_{\omega l}|^2 \delta(\omega-\omega') \delta_{ll'}\delta_{mm'},
\end{equation} 
consistently with expectations. The ray-tracing prescription \eqref{limp-} indeed neglects the portion of the wave function that never penetrates the potential barrier around the collapsing matter, which is however irrelevant to the effect of calculating the number of emitted particles \cite{Hawking1975}.

Using the ray-tracing relation \eqref{limp-} one can then retrieve the Bogolyubov coefficients by evaluating the scalar products in \eqref{Bogcont} at $\mathscr I^-$. 
For instance, proceeding as in \eqref{IBP}, we have
\begin{equation}\label{}
	\alpha_{\omega \omega' l l' m m'} = \frac{t_{\omega l}}{2\pi}\sqrt{\frac{\omega'}{\omega}}\int_{-\infty}^{v_0} dv \left(\frac{v_0-v}{C}\right)^{-\frac{i\omega}{\kappa}} e^{(\epsilon-i\omega')v}\oint d\Omega(\hat x) Y_{lm}(\hat x) Y_{l'm'}^\ast(\hat x) \,,
\end{equation}
where a small positive quantity $\epsilon$ has been introduced in order to make the integral convergent. One is thus led to
\begin{align}
	\label{solution for alpha in the hawking case}
	\alpha_{\omega\omega'll'mm'}
	&=
	\frac{t_{\omega l}}{2\pi} \sqrt{\frac{\omega'}{\omega}}\,e^{-i\omega'v_0}C^{\frac{i\omega}{\kappa}}
	\frac{\Gamma\left(1-\frac{i\omega}{\kappa}\right)}{(\epsilon-i\omega')^{1-\frac{i\omega}{\kappa}}}\delta_{ll'}\delta_{mm'}\,,\\
	\label{solution for beta in the hawking case}
	\beta_{\omega\omega'll'mm'}
	&= 
	\frac{t_{\omega l}}{2\pi} \sqrt{\frac{\omega'}{\omega}}\,e^{i\omega'v_0}C^{\frac{i\omega}{\kappa}}
	\frac{\Gamma\left(1-\frac{i\omega}{\kappa}\right)}{(\epsilon+i\omega')^{1-\frac{i\omega}{\kappa}}}\delta_{ll'}\delta_{mm'}\,.
\end{align}

Looking at the correlator \eqref{corrn} and inserting the expressions \eqref{solution for alpha in the hawking case}, \eqref{solution for beta in the hawking case} for the Bogolyubov coefficients then yields\footnote{It should be stressed once again that the geometric optics estimate \eqref{limp-} is only accurate for small $v_0-v$ and hence the Fourier transforms $\alpha_{\omega\omega' l l'm m'}$, $\beta_{\omega\omega' l l'm m'}$ are only reliable for large $\omega'$. Nevertheless, following \cite{Hawking1975}, we extend the $\omega''$ integration in \eqref{bdaggerb hawking} from $0$ to $\infty$. This should not significantly affect the leading singularity as $\omega\to\omega'$, and hence the spectral emission rate.} 
\begin{equation}  \begin{split}
	\left\langle \boldsymbol b_{\omega l m}^\dagger \boldsymbol b_{\omega' l' m'} \right\rangle 
	& = \frac{t_{\omega l}\, \overline {t_{\omega' l}}\,\delta_{ll'}\delta_{mm'}}{2\pi\sqrt{\omega \omega'}}\, C^{\frac{i}{\kappa}(\omega-\omega')} \Gamma\left(1-\tfrac{i \omega}{\kappa}\right)\Gamma\left(1+\tfrac{i \omega'}{\kappa}\right) e^{-\frac{\pi (\omega+\omega')}{2\kappa}} \\ & \times \int_0^\infty \omega''^{\frac{i}{\kappa}(\omega-\omega')}\frac{d \omega''}{2\pi\omega''}\,, \label{bdaggerb hawking}
\end{split}\end{equation}
after using
\beq \label{analytical continuatin of the log}
\log(\epsilon\pm i \omega'')=\log \omega'' \pm \frac{i\pi}{2}\,.
\eeq
The final integral gives rise to a delta function in the frequencies
as can be seen by performing the change of variable $\tau=\frac{1}{\kappa}\log \omega''$.
One is thus led to 
\begin{equation}\label{Hawking spectrum}
	\left\langle \boldsymbol b_{\omega l m}^\dagger \boldsymbol b_{\omega' l' m'} \right\rangle
	=
	\frac{|t_{\omega l}|^2\,}{e^{\frac{2\pi\omega}{\kappa}}-1}\, \delta(\omega-\omega')\delta_{ll'}\delta_{mm'}\,,
\end{equation}
after using the relation 
\begin{equation} \label{Gamma function squared}
	\left|\Gamma\left(1+ ic \right)\right|^2 = \frac{\pi c}{\sinh(\pi c)}  \,,\qquad c\in\mathbb R\,.
\end{equation}
The total number of particles emitted at each $l$, $m$ per unit frequency  $\left\langle \boldsymbol b_{\omega l m}^\dagger \boldsymbol b_{\omega l m} \right\rangle$ is infinite, since  for each $\omega$ the observer at $\mathscr I^+$ detects a steady emission of particles for all times. One can therefore conveniently turn to the spectral emission rate \cite{Page:1976df} for each $l$, $m$, obtaining\footnote{This can be seen from \eqref{Hawking spectrum} by applying the following standard replacement for a large observation time $\tau$ and $\omega$ close to  $\omega'$
	\begin{equation}
		\delta(\omega-\omega')\simeq\int_{-\frac{\tau}{2}}^{\frac{\tau}{2}}\frac{dt}{2\pi} \, e^{i(\omega-\omega')t}\rightarrow \frac{\tau}{2\pi}\,.
	\end{equation}
} 
\begin{equation}\label{Hawking rate}
	dN_{lm}(\omega)=\frac{1}{2\pi}\, \frac{|t_{\omega l}|^2}{e^{\frac{2\pi\omega}{\kappa}}-1}\,d\omega\,.
\end{equation}
We have thus retrieved the well-known Bose-Einstein distribution of massless scalar particles at temperature $T=\kappa / (2\pi)$ emitted by the black-hole, weighted by the transmission coefficient $|t_{\omega l}|^2$. Summing over angular labels yields the spectral emission rate at the frequency $\omega$,
\begin{equation}\label{dNwstandard}
	dN(\omega)=\frac{1}{2\pi}\, \frac{g(\omega)}{e^{\frac{2\pi\omega}{\kappa}}-1}\,d\omega\,,
\end{equation}
where we have defined the density of states
\begin{equation}\label{dos}
	g(\omega)=\sum_l(2l+1)|t_{\omega l}|^2\,.
\end{equation}

To obtain a different perspective on the result which will be useful in the next sections, let us introduce an alternative basis of mode functions, which we call the $\omega,\hat q$ basis, defined by
\begin{align}\label{changeofbasis}
	\psi_{\omega \hat q}(v,r,\hat x) &= \sum_{l,m}\psi_{\omega l m}(v,r,\hat x) Y^\ast_{l m}(\hat q)
\end{align}
where $\hat q$ is a unit vector. In this new basis, by denoting $\delta(\hat q, \hat x)$ the invariant delta function on the sphere, we have
\begin{align}\label{asdeltaf}
	f_{\omega \hat q} (v,r,\hat x) &\,\sim\, \frac{1}{\sqrt{2\pi\omega}\,r}\, e^{i\omega v}\,\delta(\hat q, \hat x)\qquad (\text{at }\mathscr I^-) \, ,\\
	\label{asdeltap}
	p_{\omega \hat q} (u,r,\hat x) &\,\sim\, \frac{1}{\sqrt{2\pi\omega}\, r}\, e^{i\omega u}\,\delta(\hat q, \hat x)\qquad (\text{at }\mathscr I^+)\,.
\end{align}
The advantage of this formulation is to replace the spherical harmonics with delta-functions on the sphere, thus yielding a family of states with an asymptotically well defined direction of propagation.
Using the ray-tracing formula \eqref{u(v)} for the spherically symmetric collapse, we then have
\begin{equation}\label{as-delta}
	p_{\omega \hat q} (u,r,\hat x) \,\sim\, \frac{t_\omega(\hat q, \hat x)}{\sqrt{2\pi\omega}\, r}\,e^{i\omega u(v)}\, \Theta(v_0-v)\qquad (\text{at }\mathscr I^-)\,,
\end{equation}
where the transmission amplitude in this basis has now the form
\begin{equation}\label{smearedt}
	t_{\omega}(\hat q, \hat x) = \sum_{l,m}Y_{lm}^\ast(\hat q) \, t_{\omega l} \, Y_{lm}(\hat x)
	=
	\frac{1}{4\pi}\,\sum_l (2l+1)t_{\omega l} P_l(\hat q\cdot \hat x)\,,
\end{equation}
where we used the addition theorem \eqref{YYPx} in the second step.
Note that, although $p_{\omega \hat q}(u,r,\hat x)$ is localized to a point on the celestial sphere in the far future, the dispersion induced by the gravitational field effectively smears it out in the far past according to \eqref{smearedt}.
However, in the approximation $t_{\omega l}\simeq t_\omega$ in which the transmission amplitude is only a function of the frequency, $t_{\omega}(\hat x, \hat q)\simeq t_{\omega}\, \delta(\hat x, \hat q)$ and the effects of dispersion disappear.

The Bogolyubov coefficients in this basis read
\begin{align}
	\label{alphadelta}
	\alpha_{\omega\omega'\hat q \hat q'}
	&=
	\frac{t_{\omega}(\hat q, \hat q')}{2\pi} \sqrt{\frac{\omega'}{\omega}}\,e^{-i\omega'v_0}C^{\frac{i\omega}{\kappa}}
	\frac{\Gamma\left(1-\frac{i\omega}{\kappa}\right)}{(\epsilon-i\omega')^{1-\frac{i\omega}{\kappa}}}\,,\\
	\label{betadelta}
	\beta_{\omega\omega'\hat q \hat q'}
	&= 
	\frac{t_{\omega}(\hat q, \hat q')}{2\pi} \sqrt{\frac{\omega'}{\omega}}\,e^{i\omega'v_0}C^{\frac{i\omega}{\kappa}}
	\frac{\Gamma\left(1-\frac{i\omega}{\kappa}\right)}{(\epsilon+i\omega')^{1-\frac{i\omega}{\kappa}}}\,,
\end{align}
while the Hawking spectrum takes the form
\begin{equation}\label{HawkingSpectrumdelta}
	\left\langle \boldsymbol b^\dagger_{\omega \hat q} \,\boldsymbol b_{\omega' \hat q'}  \right\rangle =  \frac{\Gamma_{\omega}(\hat q, \hat q')}{e^{\frac{2\pi \omega}{\kappa}}-1}\,\delta(\omega-\omega') \,,
\end{equation}
where $	\Gamma_{\omega}(\hat q, \hat q')$ stores the information about the transmission coefficients 
\begin{equation}\label{Gamma}
	\Gamma_{\omega}(\hat q, \hat q') = \oint d\Omega(\hat q'')\, t_\omega(\hat q, \hat q'') \overline{t_\omega} (\hat q', \hat q'')=\sum_{l,m} Y_{lm}(\hat q) |t_{\omega l}|^2 Y_{lm}^\ast(\hat q')\,,
\end{equation}
or equivalently, again thanks to the identity \eqref{YYPx},
\begin{equation}\label{Gamma2}
	\Gamma_{\omega}(\hat q, \hat q') = \frac{1}{4\pi}\sum_l (2l+1) |t_{\omega l}|^2 P_l(\hat q\cdot \hat q')\,.
\end{equation}
The spectral emission rate per unit solid angle is then given by
\begin{equation}\label{Hawkingemission}
	dN(\omega, \hat q) = \frac{1}{8\pi^2}\, \frac{ g(\omega) }{e^{\frac{2\pi\omega}{\kappa}}-1}\, d\omega\,d\Omega(\hat q) \,.
\end{equation}

\section{BMS Group and Finite Transformations} 
\label{sec:BMSreview}

In this section we provide a brief summary of the asymptotic symmetries of asymptotically flat spacetimes, also known as  BMS group \cite{Bondi:1962px,Sachs:1962wk,Sachs:1962zza}.

Let us recall the main steps leading to Penrose's conformal compactification \cite{Penrose:1965am,Penrose:1964ge} (see \cite{Geroch:1977jn,Wald:1984rg,Ashtekar:1987tt} for excellent introductory presentations) and to the construction of future null infinity $\IP$ for Minkowski spacetime, whose metric written in retarded coordinates takes the form
\begin{equation}\label{}
ds^2=g_{ab}\,dx^{a}dx^{b}=-du^2 - 2 du\, dr + r^2 \gamma_{AB}\,d\xi^{A} d\xi^{B}\,.
\end{equation}
Here $u=t-r$ is the retarded time coordinate, $\xi^{A}$ are two angles and $\gamma_{AB}$ denotes the metric on the unit sphere.
Letting $\Omega=\frac{1}{r}$ and multiplying $ds^2$ by the conformal factor $\Omega^2$, one obtains the metric
\begin{equation}\label{}
d\tilde s^2= \tilde g_{ab}\,dx^{a}dx^{b}=-\Omega^2 du^2 + 2 du\, d\Omega + \gamma_{AB}\,d\xi^{A} d\xi^{B}\,.
\end{equation}  
The meaning of this step is to bring ``infinity'', associated to $r=\infty$ or $\Omega=0$, to a finite distance.
Restricting indeed to a surface $\Sigma_\Omega$ of constant $\Omega$, which corresponds to a surface of fixed radius in the original coordinates,
\begin{equation}\label{}
d\tilde s^2\big|_{\Sigma_\Omega} =-\Omega^2 du^2+  {\gamma_{AB}}\,d\xi^{A} d\xi^{B}\,,
\end{equation}
and considering in particular the boundary surface at $\Omega=0$, that is, future null infinity $\mathscr I^+$, one finds 
\begin{equation}\label{scrimetric}
d\tilde s^2\big|_{\mathscr I^+} =0 \cdot du^2 + {\gamma_{AB}}\,d\xi^{A} d\xi^{B}\,,
\end{equation}
which is the metric induced on $\mathscr I^+$. The coefficient $0$ in front of $du^2$ has been made apparent to underline that $u$ is indeed a null coordinate on $\mathscr I^+$. 

The normal co-vector to the surface $\Sigma_\Omega$ is given by $\partial_a\Omega$ and one can find the associated vector by applying the inverse metric,  
\begin{equation}\label{}
\tilde n^{a}=\tilde g^{ab}\partial_b\Omega\,,\qquad
\tilde n^{a}\partial_a=\partial_u + \Omega^2\partial_\Omega\,.
\end{equation}
On $\mathscr I^+$, where $\Omega=0$, this reads
\begin{equation}\label{scrin}
\tilde n\big|_{\mathscr I^+}=\partial_u\,.
\end{equation}
This normal vector and the metric \eqref{scrimetric} comprise the two main geometric structures characterizing $\mathscr I^+$.

The choice of multiplying the original metric by $\Omega^2=\frac{1}{r^2}$ is somewhat arbitrary. Indeed, an equivalent characterization of the boundary $\mathscr I^+$ ought to obtain by choosing a different conformal factor $\omega^2\Omega^2$
in the above discussion, provided that $\omega$ is a smooth function which is finite and nonzero near the boundary. The new unphysical metric and its inverse are then
effectively obtained by the replacements
\begin{equation}
\tilde g_{ab}\to \omega^2 \tilde g_{ab}\,,\qquad
\tilde g^{ab}\to \frac{1}{\omega^2} \tilde g^{ab}
\end{equation}
and
\begin{equation}\label{scrimetric'}
d\tilde s^2\big|_{\mathscr I^+} = 0\cdot du^2+ {\omega^2}\,{\gamma_{AB}}\,d\xi^{A} d\xi^{B}\,.
\end{equation}
The relevant normal vector instead becomes
\begin{equation}\label{}
\tilde n^{a}=\frac{1}{\omega^2}\,\tilde g^{ab}\partial_b(\omega\Omega)
\end{equation}
so that on $\mathscr I^+$
\begin{equation}\label{scrin'}
\tilde n\big|_{\mathscr I^+}=\frac{1}{\omega}\,\partial_u\,.
\end{equation}

The BMS group is the set of transformations that preserve $\mathscr I^+$ and map its geometric structure to itself up to a conformal factor $\omega$  according to
\begin{equation}\label{BMSconditions}
\partial_u \to \frac{1}{\omega}\partial_u\,,\qquad
\gamma_{AB}\to \omega^2 \gamma_{AB}\,,
\end{equation}
as dictated by the comparison between \eqref{scrin} and \eqref{scrin'} as well as between \eqref{scrimetric} and \eqref{scrimetric'}. This equivalence class of line elements and normal vectors on $\mathscr I^+$ defines the so-called universal structure of asymptotically flat spacetimes. By a standard result, any spacetime satisfying the BMS fall-off conditions, which characterize the notion of asymptotic flatness in the physical space by specifying how the metric reduces to the flat one asymptotically far from matter sources, admits indeed a conformal boundary in the unphysical space whose properties fall within the universal structure, and vice versa \cite{Ashtekar:1987tt,Wald:1984rg}.

Consider now a generic transformation given by $u'=F(u,\xi)$ and $\xi'^{A}=G^{A}(u,\xi)$, and impose \eqref{BMSconditions}. From
\begin{equation}\label{}
\partial_u=\partial_uF\,\partial_u'+\partial_u G^{A}\,\partial'_A = \omega \partial_u'
\end{equation}
we read off
\begin{equation}\label{}
\partial_uF=\omega\,,\qquad
\partial_u G^{A}=0\,,
\end{equation}
so that the $u$-dependence of $G^{A}$ can be dropped, $G^{A}(u,\xi)=G^{A}(\xi)$.
Furthermore, from
\begin{equation}\label{}
\gamma_{AB}(\xi)=\frac{1}{\omega^2(u,\xi)}\partial_A G^{C}(\xi) \gamma_{CD}(\xi') \partial_B G^{D}(\xi)
\end{equation}
we see that $\omega$ must be $u$-independent,
\begin{equation}\label{}
\omega(u,\xi)=R(\xi),
\end{equation} 
and thus\begin{equation}\label{}
F(u,\xi)=T(\xi)+u\,R(\xi)\,,
\end{equation}
where $T(\xi)$ is a generic angular function.
To summarize, a finite BMS transformation is specified by an arbitrary function $T(\xi)$ of the angles and by a conformal transformation $G^{A}(\xi)$  of the unit sphere with conformal factor $R(\xi)$ according to
\begin{equation}\label{}
u'=T(\xi)+u\,R(\xi)\,,\qquad
\xi'^{A}=G^{A}(\xi)\,,\qquad
\gamma_{AB}(\xi)=\frac{1}{R^2(\xi)}\partial_A G^{C}(\xi) \gamma_{CD}(\xi') \partial_B G^{D}(\xi)\,.
\end{equation}

Supertranslations are obtained when $G^{A}$ is the identity (and $R(\xi)=1$),
\begin{equation}\label{ST}
u'=u+T(\xi)\,,\qquad \xi'=\xi\,.
\end{equation}
They form an infinite-dimensional normal subgroup of the BMS group and contain the standard spacetime translations
as their unique normal subgroup of dimension four \cite{Sachs:1962wk}.
For $T=0$ one is left with the transformations
\begin{equation}\label{SR}
u'=u\,R(\xi)\,,\qquad
\xi'^{A}=G^{A}(\xi)\,,\qquad
\gamma_{AB}(\xi)=\frac{1}{R^2(\xi)}\partial_A G^{C}(\xi) \gamma_{CD}(\xi') \partial_B G^{D}(\xi)\,,
\end{equation}
parametrized by conformal mappings of the sphere to itself (see Appendix \ref{app:conformal}).
These admit the Lorentz group $SL(2,\mathbb C)\simeq SO(1,3)$ as the unique subfamily of globally-well-defined maps. However, one may also retain more general transformations satisfying \eqref{SR} everywhere except at localized singularities, such as poles or branch cuts, thus enhancing standard boosts and rotations to \emph{superrotations} \cite{Barnich:2009se,Barnich:2010eb,Barnich:2011ct}. These generalized Lorentz transformations will be the main source of novelties in the ensuing discussion on black hole spectra.

At the infinitesimal level, this extension corresponds to the familiar  enhancement from the six-dimensional space of globally well-defined conformal Killing vector to two copies of the infinite-dimensional Witt algebra, or its central extension the Virasoro algebra, with manifold applications in two-dimensional conformal field theory. In fact, it has been advocated that, by suitably relaxing the BMS boundary conditions, infinitesimal superrotations should be even further extended to arbitrary diffeomorphisms on the sphere, a proposal which opens the way for even more powerful asymptotic symmetries \cite{Campiglia:2015yka}.

The construction illustrated above for future null infinity can be repeated for past null infinity, leading to another copy of the BMS group acting on $\mathscr I^-$ and in principle independent of the one defined on $\mathscr I^+$. 
These two BMS actions should however be appropriately linked to one another in order to identify a symmetry for the $S$-matrix of massless states on asymptotically flat spacetimes. 
Indeed the conservation of BMS charges at spatial infinity and the relation between BMS symmetries and soft theorems for scattering amplitudes point to a link between past and future null infinity in the form of an \emph{antipodal matching} condition \cite{Strominger2014,He2015,Strominger:2017zoo,Hawking2017}. 
Considering the future boundary $\mathscr I^-_+$ of $\mathscr I^-$ and the past boundary $\mathscr I^+_-$ of $\mathscr I^+$, this condition requires that BMS transformations on $\mathscr I^-$ be matched to those on $\mathscr I^+$ by requiring that their actions agree on $\mathscr I^-_+\simeq\mathscr I^+_-$, after identifying  antipodal points on these two spheres. 
On top of these considerations, the antipodal matching can be seen to hold for a rather general class of solutions of the field equations and also affords a very natural geometric interpretation in Penrose's conformal picture \cite{Strominger:2017zoo}.

For these reasons, following \cite{Strominger:2017zoo,Hawking2017}, the antipodal matching is implicitly built in our choice of angular coordinates at $\mathscr I^-$ and $\mathscr I^+$ throughout the rest of the paper, meaning that $\xi$ at $\mathscr I^-$ shall denote the antipodal point of $\xi$ at $\mathscr I^+$. However, as we shall see, the spectrum eventually turns out to only depend on the copy of BMS acting at $\mathscr I^+$, both for supertranslations and for superrotations, thus making the antipodal identification immaterial as far as the present discussion is concerned.

\section{Supertranslations}
\label{sec: supertranslations}

Having now at hand the action of supertranslations, let us review how Hawking's calculation is modified by these transformations, a problem also discussed in \cite{Javadinezhad2019,Compere2019}. As already pointed out by Hawking in his seminal paper \cite{Hawking1975} the thermal emission rate is eventually unaffected by supertranslations. However, we also find non-trivial correlations for different $l,m$ modes which have been overlooked in previous work. 

The asymptotics of supertranslated states differ from those of a spherically symmetric collapse \eqref{limp+}, \eqref{limf-} by an angle-dependent shift of the retarded and advanced time coordinates,  
\begin{align} \label{supertranslation expansion P}
	p_{\omega l m} (u,r,\xi)&\,\sim\,  \frac{1}{\sqrt{2\pi\omega}\, r}\, e^{i\omega \left(u-T^{+}(\xi)\right)} Y_{lm}(\xi)\qquad (\text{at }\mathscr I^+) \\ 
	\label{supertranslation expansion F}
	f_{\omega l m} (v,r,\xi)&\,\sim\,  \frac{1}{\sqrt{2\pi\omega}\, r}\, e^{i\omega \left(v-T^{-}(\xi)\right)} Y_{lm}(\xi)\qquad (\text{at }\mathscr I^-)\,,
\end{align}
where $T^{\pm}(\xi)$ characterize supertranslations at ${\mathscr I}^\pm$ as in eq.~\eqref{ST}. Since we focus here on supertranslated states, we omit for simplicity the superscript ``ST'' and write e.g. $p_{\omega l m}$ instead of $p^\text{ST}_{\omega l m}$, believing that no confusion should arise.
Note that, despite the modified angular dependence, such functions are still orthonormal with respect to the standard scalar product. Furthermore, they still give rise to asymptotic solutions of the Klein--Gordon equation, since they differ from the standard ones by an asymptotic symmetry.

\subsection{The $\omega,\hat q$ basis}
In the case of supertranslated states, because the functions $T^{\pm}$ affect the angular dependence of the solutions, it is convenient to turn to the $\omega,\hat q$ basis \eqref{changeofbasis}, where the supertranslated states obey the simple asymptotics
\begin{align}\label{asdeltaST}
		p_{\omega \hat q} (u,r,\hat x) &\,\sim\, \frac{e^{-i\omega T^+(\hat q)}}{\sqrt{2\pi\omega}\,r}\, e^{i\omega u}\,\delta(\hat q, \hat x)\qquad (\text{at }\mathscr I^+) \\
	f_{\omega \hat q} (v,r,\hat x) &\,\sim\, \frac{e^{-i\omega T^-(\hat q)}}{\sqrt{2\pi\omega}\, r}\, e^{i\omega v}\,\delta(\hat q, \hat x)\qquad (\text{at }\mathscr I^-) \, .
\end{align}
Therefore, the corresponding limit of $p_{\omega\hat q}$ at $\mathscr I^-$ coincides with the standard one \eqref{as-delta} up to an overall phase, 
\begin{equation}\label{as-deltaST}
	p_{\omega \hat q} (u,r,\hat x) \,\sim\, t_\omega(\hat q, \hat x) e^{-i\omega T^+(\hat q)} \frac{ e^{i\omega u(v)}}{\sqrt{2\pi\omega}\, r}\,  \Theta(v_0-v)\qquad (\text{at }\mathscr I^-)\,.
\end{equation}
The Bogolyubov coefficients for the supertranslated case then read 
\begin{align}
	\label{alphaHdeltaST}
	\alpha_{\omega\omega'\hat q \hat q'}
	&=
	\frac{t_{\omega}(\hat q, \hat q')}{2\pi}\, \sqrt{\frac{\omega'}{\omega}}\,e^{-i\omega'v_0}C^{\frac{i\omega}{\kappa}}
	\frac{\Gamma\left(1-\frac{i\omega}{\kappa}\right)}{(\epsilon-i\omega')^{1-\frac{i\omega}{\kappa}}}\, e^{-i\omega T^+(\hat q)+i\omega' T^-(\hat q')}\,,\\
	\label{betaHdeltaST}
	\beta_{\omega\omega'\hat q \hat q'}
	&= 
	\frac{t_{\omega}(\hat q, \hat q')}{2\pi}\, \sqrt{\frac{\omega'}{\omega}}\,e^{i\omega'v_0}C^{\frac{i\omega}{\kappa}}
	\frac{\Gamma\left(1-\frac{i\omega}{\kappa}\right)}{(\epsilon+i\omega')^{1-\frac{i\omega}{\kappa}}}\, e^{-i\omega T^+(\hat q)-i\omega' T^-(\hat q')}\,,
\end{align}
while the spectrum can be read off from
\begin{equation}\label{mainST}
	\left\langle \boldsymbol b^\dagger_{\omega \hat q} \,\boldsymbol b_{\omega' \hat q'}  \right\rangle =  \frac{\Gamma_{\omega}(\hat q, \hat q')}{e^{\frac{2\pi \omega}{\kappa}}-1}\,\delta(\omega-\omega') \, e^{-i\omega \left[T(\hat q)-T(\hat q')\right]}\,,
\end{equation}
with $\Gamma(\hat q, \hat q')$ as in \eqref{Gamma}. The last equation is the first main result of this work. It shows that the two-point correlator is modified, compared to Hawking's result, by an off-diagonal phase while the spectral emission rate at fixed angle, associated with the diagonal entries, coincides with the standard one
\eqref{Hawkingemission}. This off-diagonal part has not be pointed out in previous works either because the angular dependence of the transmission amplitude has been neglected \cite{Javadinezhad2019} or because a factorized form of the Bogolyubov coefficients was assumed \cite{Compere2019}. In fact, in the $l,m$ basis the final result also depends on $T^+$ but in a more cumbersome way as we will show in the next section,.
 
The off-diagonal phases disappear in the approximation that the transmission coefficient is a function only of $\omega$, $|t_{\omega l}|^2
\simeq |t_\omega|^2$, in which case $\Gamma_\omega(\hat q, \hat q')\simeq |t_\omega|^2 \delta(\hat q, \hat q')$ and one retrieves Hawking's result. This approximation is accurate for very large frequencies but it breaks down for sufficiently large angular momentum, $l \gtrsim \omega/\kappa$,  even in the high-energy limit  $\omega/\kappa \gg 1$, due to the large potential barrier surrounding the black hole which makes the transmission less probable.
Keeping the $l$-dependence in $t_{\omega l}$ is also crucial to retrieve a finite number rate and power spectrum as we will discuss in Section \ref{sec: power spectrum}.

\subsection{The $\omega,l,m$ basis}

For completeness, and for the reader more familiar with the  conventional $\omega,l,m$ basis, we provide here the same derivation in this basis.

To find the spectrum one can follow the same steps as in Section \ref{sec: Hawking spectrum} with minor modifications. 
The basic ingredient that we need is the limit of $p_{\omega l m}$ at $\IM$, i.e. the generalization of \eqref{limp-} to the supertranslated case. To obtain it we note that \eqref{supertranslation expansion P} can be written as follows
\begin{equation}\label{pSTdecomp}
	p_{\omega l m} (u,r,\xi)\,\sim\, \sum_{l'm'} c^{(+)}_{\omega ll'mm'}
	\left[\frac{1}{\sqrt{2\pi\omega}\, r}\, e^{i\omega u} \,Y_{l'm'}(\xi)\right]
	\qquad (\text{at }\mathscr I^+)\,,
\end{equation}
where we defined
\begin{equation}\label{}
	c^{(\pm)}_{\omega ll'mm'} = \int d\Omega(\xi) Y_{lm}(\xi) e^{-i\omega T^\pm(\xi)} Y^\ast_{l'm'}(\xi)\,.
\end{equation}
Therefore, employing the ray-tracing relation \eqref{limp-} for each term in the decomposition \eqref{pSTdecomp} and linearly superposing the results, we have
\begin{equation}\label{}
	p_{\omega l m} (u,r,\xi)\,\sim\, \sum_{l'm'} c^{(+)}_{\omega ll'mm'}
	\left[\frac{t_{\omega l'}}{\sqrt{2\pi\omega}\, r}\, e^{i\omega u(v)} \,Y_{l'm'}(\xi) \Theta(v_0-v)\right]
	\qquad (\text{at }\mathscr I^-)\,.
\end{equation}
The Bogolyubov coefficients are then closely related to the ones discussed in Section~\ref{sec: Hawking spectrum}, 
\begin{align}\label{alphaST}
	\alpha_{\omega\omega'l l' m m'} &= d_{\omega \omega' l l' m m'}  \frac{1}{2\pi} \sqrt{\frac{\omega'}{\omega}}\,e^{-i\omega'v_0}C^{\frac{i\omega}{\kappa}}
	\frac{\Gamma\left(1-\frac{i\omega}{\kappa}\right)}{(\epsilon-i\omega')^{1-\frac{i\omega}{\kappa}}}\,,
	\\
	\label{betaST}
	\beta_{\omega\omega'l l' m m'} &= d_{\omega (-\omega') l l' m m'} 
	\frac{1}{2\pi} \sqrt{\frac{\omega'}{\omega}}\,e^{i\omega'v_0}C^{\frac{i\omega}{\kappa}}
	\frac{\Gamma\left(1-\frac{i\omega}{\kappa}\right)}{(\epsilon+i\omega')^{1-\frac{i\omega}{\kappa}}}\,,
\end{align}
up to the overall angular mixing factor
\begin{equation}\label{}
	d_{\omega \omega' l l' m m'}
	=
	\sum_{l'' m''} 
	c^{(+)}_{\omega ll''mm''} \, t_{\omega l''} \, \overline{c^{(-)}_{\omega' l'l''m'm''}}\,.
\end{equation}
The two-point correlator then reads
\begin{equation}\label{bbST1}
	\left\langle \boldsymbol b_{\omega l m}^\dagger \boldsymbol b_{\omega' l' m'} \right\rangle
	=
\, \frac{	\Delta_{\omega ll'mm'}  }{e^{\frac{2\pi\omega}{\kappa}}-1} \delta(\omega-\omega')
\end{equation}
with
\begin{equation}\label{bbST2}
	\Delta_{\omega l l' m m'}
	=
	\sum_{l'' m''} 
	c^{(+)}_{\omega ll''mm''} \, |t_{\omega l''}|^2 \, \overline{c^{(+)}_{\omega l'l''m'm''}}\,.
\end{equation}
In this basis, the correlation between different $l,m$ modes looks more involved precisely because of the asphericity of the supertranslation functions.
Note that eq. \eqref{bbST1} differs from the one obtained in \cite{Javadinezhad2019,Compere2019}. The two results would agree under the approximation $t_{\omega l}\simeq t_\omega$ in which the transmission amplitude is only a function of the frequency. In that case the spectrum again simplifies and we recover Hawking's result
\begin{equation}\label{}
	\Delta_{\omega l l' m m'} \simeq |t_\omega|^2 \delta_{ll'} \delta_{mm'}\,.
\end{equation}
However, as mentioned above the approximation breaks down for large $l$ or when computing quantities which involve the summation over $l$ modes.
Note also that \eqref{bbST1} could also have been directly obtained from \eqref{mainST} by just converting each $\boldsymbol b_{\omega \hat q}$ to the $\omega,l,m$ basis by means of eq. \eqref{changeofbasis}.

Regarding the the spectral emission rate, the relevant matrix elements are the diagonal ones, $l=l'$, $m=m'$, which can be made more explicit using the addition theorem \eqref{YYPx},
\begin{equation}\label{}
	\Delta_{\omega l l m m}
	=
	\sum_{l'} \frac{2l'+1}{4\pi} |t_{\omega l'}|^2
	\int d\Omega(\xi)
	\int d\Omega(\xi') Y_{lm}(\xi) e^{-i\omega [T^+(\xi)-T^+(\xi')]} 
	Y^\ast_{lm}(\xi') 
	P_{l'}(\cos\theta_{\xi, \xi'})
	\,,
\end{equation}
so that
\begin{equation}\label{}
	dN_{l m}(\omega)
	=
	\frac{1}{2\pi} \, \frac{\Delta_{\omega ll mm}}{e^{\frac{2\pi\omega}{\kappa}}-1}\, d\omega\,.
\end{equation}
After summing over $l$, $m$ the result then simplifies to the standard one \eqref{dNwstandard}.

\section{Boosts and Superrotations}
\label{sec: superrotations}

We now move to the more interesting case of boosts and superrotations. The derivation will be valid for both types of transformations but, for simplicity, we will refer to both as superrotated states and only distinguish between the two when necessary. 

As discussed in Section~\ref{sec:BMSreview}, superrotations are not globally well defined as they introduce singularities on the celestial sphere. In particular we will have to deal with singularities at the north and south pole where  the conformal factors $R_\pm \to 0$. However, as we will discuss, at the level of the two-point function the angular integral involved in the calculation does not give rise to divergences for any $l$, $m$.

Starting from the expansions \eqref{limp+} and \eqref{limf-}, after performing the finite transformations in eq.~\eqref{SR} the superrotated states take the asymptotic form
\begin{align} \label{superrotation expansion P}
	p_{\omega l m} (u,r,\xi)&\,\sim\,  \frac{R_+(\xi)}{\sqrt{2\pi\omega}\, r}\, e^{i\omega u R_+(\xi)} Y_{lm}(\xi_+)\qquad (\text{at }\mathscr I^+) \\ 
	\label{superrotation expansion F}
	f_{\omega l m} (v,r,\xi)&\,\sim\,  \frac{R_-(\xi)}{\sqrt{2\pi\omega}\, r}\, e^{i\omega v R_-(\xi)} Y_{lm}(\xi_-)\qquad (\text{at }\mathscr I^-)\,,
\end{align}
where $\xi = (\theta, \phi)$ and 
\begin{equation}
	\xi_\pm =  G_\pm(\xi)\,
\end{equation}
denote the new angular coordinates at $\IP$ and $\IM$.
We once again omit the superscript ``SR'', writing e.g. $p_{\omega l m}$ instead of $p^\text{SR}_{\omega l m}$ for ease of notation.
Note that, together with the transformations dictated by \eqref{SR}, an overall factor of $R_\pm$ has been included in order to account for the transformation $r\to r/R_\pm(\xi)$ to leading order.  Equivalently, one may note that the asymptotic fields must have unit conformal weight at null infinity.

Let us first check whether the states \eqref{superrotation expansion P} and \eqref{superrotation expansion F} satisfy the proper orthonormality conditions, considering for instance
\begin{equation}\label{}
	(f_{\omega lm}, f_{\omega'l'm'}) =
	\frac{\omega'+\omega'}{2\sqrt{\omega \omega'}} \int \frac{dv}{2\pi}\,  e^{iv(\omega-\omega')R_-(\xi)}\int d\Omega(\xi) R_-(\xi)^3 Y_{lm}(\xi_-)Y^\ast_{l'm'}(\xi_-)\,,
\end{equation}
Here the integral in $v$ gives rise to $\delta((\omega-\omega')R_-(\xi))$, whose support is located at $\omega=\omega'$ or where $R_-(\xi)$ has zeros. In Appendix \ref{app:R=0} we explain why the extra contributions due to the zeros of $R_-$ vanish, effectively justifying
\begin{equation}
	\delta\left((\omega-\omega')R_-(\xi)\right)= \frac{1}{R_-(\xi)}\, \delta(\omega-\omega')\, . 
\end{equation}
Using this identity together with 
\eqref{confintegral} to the conformal transformation $\xi_-$ yields
\begin{equation}\label{}
	(f_{\omega lm}, f_{\omega'l'm'}) =\delta(\omega-\omega')\delta_{ll'} \delta_{mm'}
\end{equation}
provided that $G_-(\xi)$ is a one-to-one mapping of the sphere onto itself. As detailed below, we will be interested in more general transformations that can wrap $n$ times around the celestial sphere along the azimuthal direction.
The superrotated spherical harmonics are then assumed to satisfy the following orthogonality and overcomplenetess relations
\begin{equation}\label{}
	\int d\Omega(\xi_\pm) Y_{lm}(\xi_\pm)Y^\ast_{l'm'}(\xi_\pm)= n\, \delta_{ll'} \delta_{mm'}\,,\qquad
	\frac{1}{n}\sum_{lm} Y_{lm}(\xi_\pm) Y^\ast_{lm}(\xi_\pm') = \delta(\xi_\pm, \xi_\pm')\,,
\end{equation}
thus yielding
\begin{equation}\label{nnorm}
	\left(f_{\omega l m},f_{\omega',l',m'}\right)=n\,\delta(\omega-\omega')\delta_{ll'} \delta_{mm'}\,.
\end{equation}
We will come back to this point when considering a specific class of superrotations for which these properties can be checked explicitly. 

Now we proceed to the computation of the spectrum. Similarly to the case of supertranslations, we first discuss the derivation in the $\omega,\hat q$ basis, where the presentation is more transparent, and later give the derivation in the $\omega, l , m$ basis. 

\subsection{The $\omega,\hat q$ basis}

In the basis given by \eqref{changeofbasis} the asymptotic limits of the mode functions  read
\begin{align}\label{asdeltafSR}
	f_{\omega \hat q} (v,r,\hat x) &\,\sim\, \frac{R_-(\hat x)}{\sqrt{2\pi\omega}\,r}\, e^{i\omega v R_-(\hat x)}\,\delta(\hat q, \hat x_-)\qquad (\text{at }\mathscr I^-)\\
	\label{asdeltapSR}
	p_{\omega \hat q} (u,r,\hat x) &\,\sim\, \frac{R_+(\hat x)}{\sqrt{2\pi\omega}\,r}\, e^{i\omega u R_+(\hat x)}\,\delta(\hat q, \hat x_+)\qquad (\text{at }\mathscr I^+)\,,
\end{align}
where the unit vector
\begin{equation}
	\hat x_\pm = G_\pm (\hat x)
\end{equation}
denotes conformal transformations with conformal factors $R_\pm(\hat x)$. 
Using the identity \eqref{confintegral} together with the properties of the Dirac delta function, we can write these asymptotic conditions in the form
\begin{align}\label{asdeltafSRd}
	f_{\omega \hat q} (v,r,\hat x) &\,\sim\, \frac{e^{i\omega v R_-(\hat Q_-)}}{\sqrt{2\pi\omega}\,r} \,\frac{\delta( \hat Q_-,\hat x)}{R_-(\hat Q_-)} \qquad (\text{at }\mathscr I^-)\\
	\label{asdeltapSRd}
	p_{\omega \hat q} (u,r,\hat x) &\,\sim\, \frac{e^{i\omega v R_+(\hat Q_+)}}{\sqrt{2\pi\omega}\,r}\, \frac{\delta( \hat Q_+,\hat x)}{R_+(\hat Q_+)}\qquad (\text{at }\mathscr I^+)\,,
\end{align}
where $\hat Q_\pm$ are defined as the solutions of 
\begin{equation}
	G_{\pm}(\hat Q_\pm)=\hat q\,.
\end{equation}
Such solutions are unique for boosts, while there will be $n$ possible values of $\hat Q_\pm$ for superrotations. In the latter case, a sum over all solutions is understood in \eqref{asdeltafSRd} and \eqref{asdeltapSRd}.

From \eqref{asdeltapSRd} it is again possible to deduce the correct asymptotics of $p_{\omega \hat q}$ in the far past by linearity using \eqref{as-delta}, obtaining
\begin{equation}\label{}
	p_{\omega \hat q} (u,r,\hat x) \,\sim\, \frac{t_{\omega_+}(\hat Q_+,\hat x)}{\sqrt{2\pi\omega}\,r}\, \frac{e^{i\omega_+ u(v)}}{R_+(\hat Q_+)} \, \Theta(v_0-v)\,\qquad (\text{at }\mathscr I^-)\,,
\end{equation}
where we defined
\begin{equation}\label{w+}
	\omega_\pm = \omega R_\pm(\hat Q_\pm)\,.
\end{equation}
The Bogolyubov coefficients can be determined evaluating the scalar products $(p_{\omega \hat q},f_{\omega' \hat q'})$, $(p_{\omega \hat q},\bar f_{\omega' \hat q '})$, yielding the rather compact expressions
\begin{align}
	\label{alphadeltaSR}
	\alpha_{\omega\omega'\hat q \hat q'}
	&=
	\frac{t_{\omega_+}(\hat Q_+,\hat Q'_-)}{2\pi n \, R_+(\hat Q_+)} \sqrt{\frac{\omega'}{\omega}}\,e^{-i\omega'_-v_0}C^{\frac{i\omega_+}{\kappa}}
	\frac{\Gamma\left(1-\frac{i\omega_+}{\kappa}\right)}{(\epsilon-i\omega'_-)^{1-\frac{i\omega_+}{\kappa}}}\,,\\
	\label{betadeltaSR}
	\beta_{\omega\omega'\hat q \hat q'}
	&=
	\frac{t_{\omega_+}(\hat Q_+,\hat Q'_-)}{2\pi n\, R_+(\hat Q_+)} \sqrt{\frac{\omega'}{\omega}}\,e^{i\omega'_-v_0}C^{\frac{i\omega_+}{\kappa}}
	\frac{\Gamma\left(1-\frac{i\omega_+}{\kappa}\right)}{(\epsilon+i\omega'_-)^{1-\frac{i\omega_+}{\kappa}}}\,,
\end{align}
where analogously $\omega'_\pm = \omega' R_\pm(\hat Q_-')\,$.
The two-point function must be instead recovered from
\begin{equation}\label{}
	\left\langle \boldsymbol b^\dagger_{\omega \hat q} \,\boldsymbol b_{\omega' \hat q'}  \right\rangle
	=
	\int_0^\infty d\omega'' \oint d\Omega(\hat q'') \,
	\beta_{\omega\omega''\hat q \hat q''}
	\bar\beta_{\omega'\omega''\hat q' \hat q''}\,.
\end{equation}
To evaluate this integral, it is convenient to first let $\lambda = \omega'' R_-(\hat Q''_-)$, so that
\begin{equation}\label{}
	\begin{aligned}
		\left\langle \boldsymbol b^\dagger_{\omega \hat q} \,\boldsymbol b_{\omega' \hat q'}  \right\rangle
		=\ &
		\int_0^\infty \frac{d\lambda\, \lambda}{(2\pi n)^2\sqrt{\omega \omega'}}
		\frac{C^{\frac{i(\omega_+-\omega'_+)}{\kappa}}}{R_+(\hat Q_+) R_+(\hat Q_+')}
		\frac{\Gamma\left(1-\frac{i\omega_+}{\kappa}\right)\Gamma\left(1+\frac{i\omega'_+}{\kappa}\right)}{(\epsilon+i\lambda)^{1-\frac{i\omega_+}{\kappa}}(\epsilon-i\lambda)^{1+\frac{i\omega'_+}{\kappa}}}\\
		&\times \oint d\Omega(\hat Q''_-)\,  t_{\omega_+}(\hat Q_+,\hat Q''_-) \overline{t_{\omega'_+}}(\hat Q'_+,\hat Q''_-)\,,
	\end{aligned}
\end{equation}
where we have used \eqref{confintegral} (from right to left) to suitably rewrite the angular integral in the last line. Performing the integral with respect to $\lambda$ as in the Hawking case, with the help of \eqref{analytical continuatin of the log} and \eqref{Gamma2}, leads to
\begin{equation}\label{bbSR qbasis}
	\left\langle \boldsymbol b^\dagger_{\omega \hat q} \,\boldsymbol b_{\omega' \hat q'}  \right\rangle
	=
	\frac{\delta(\omega_+-\omega_+')}{e^{\frac{2\pi\omega_+}{\kappa}}-1}\,\frac{\Gamma_{\omega_+}(\hat Q_+, \hat Q_+')}{n\sqrt{R_+(\hat Q_+)R_+(\hat Q_+')}}\, .
\end{equation}
This is the second main result of this work. Although it might look cumbersome, it is in fact rather simple. The result shows that when allowing for superrotated states at future and past null infinity we recover Hawking's result \eqref{HawkingSpectrumdelta} but where the angles have been transformed and the frequency has been rescaled by the superrotation at future infinity. 
Namely, the previous result and \eqref{HawkingSpectrumdelta} are simply related by
\begin{equation} \label{forconclusions}
		\left\langle \boldsymbol b^\dagger_{\omega \hat q} \,\boldsymbol b_{\omega' \hat q'}  \right\rangle = \frac{\left\langle \boldsymbol b^\dagger_{\omega_+ \hat Q_+} \,\boldsymbol b_{\omega'_+ \hat Q'_+}  \right\rangle_\text{Hawking}}{n\sqrt{R_+(\hat Q_+)R_+(\hat Q_+')}}	
\end{equation}
where the factor of $n$ comes from the fact that superrotations wrap $n$ times the celestial sphere and the factors of $R_+^{-1/2}$ are due to the field's weight under conformal transformations. Note also that, similarly to supertranslations, the dependence on the superrotation at $\IM$ disappears.

Restricting to the same direction $\hat q = \hat q'$, we find
\begin{equation}\label{}
	\left\langle \boldsymbol b^\dagger_{\omega \hat q} \,\boldsymbol b_{\omega' \hat q}  \right\rangle
	=
	\frac{1}{4\pi}
	\frac{\delta(\omega-\omega')}{e^{\frac{2\pi\omega_+}{\kappa}}-1}\,\frac{g(\omega_+)}{R_+^2(\hat Q_+)}\,,
\end{equation}
which yields the spectral emission rate at fixed angle 
\begin{equation}\label{dNSR}
	dN(\omega,\hat q) = \frac{1}{8\pi^2}\,\frac{g(\omega_+)}{e^{\frac{2\pi\omega_+}{\kappa}}-1} \frac{1}{R_+^2(\hat Q_+)}\,d\omega\,   d\Omega(\hat q) \,,
\end{equation}
and, after integrating over the solid angle,
\begin{equation}\label{dNtotSR}
	dN(\omega) = \frac{1}{8\pi^2}\, d\omega\int d\Omega(\hat q)\,  \frac{g(\omega R_+(\hat q))}{e^{\frac{2\pi \omega R_+(\hat q)}{\kappa}}-1}\,.
\end{equation}

\subsection{The $\omega,l,m$ basis}

Let us now present the same derivation in the $\omega,l,m$ basis. 
In order to identify the proper limit of $p_{\omega l m}$ at $\mathscr I^-$, we proceed as in Section \ref{sec: supertranslations} and linearly decompose the superrotated state in terms of the standard ones,
\begin{equation}\label{}
	p_{\omega l m} \sim \int_0^{\infty} d\omega' \sum_{l'm'} c^{(+)}_{\omega \omega' l l' m m'}\left[\frac{1}{\sqrt{2\pi\omega'}\, r}\, e^{i\omega' u} \,Y_{l'm'}(\xi)\right]
	\qquad (\text{at }\mathscr I^+)\,,
\end{equation} 
where 
\begin{equation}\label{}
	c^{(+)}_{\omega \omega' l l' m m'} = \int d\Omega(\xi) \delta(\omega'-\omega R_+(\xi))  R_+^{\frac{3}{2}}(\xi) Y_{lm}(\xi_+) Y^\ast_{l'm'}(\xi)\,.
\end{equation}
Note that in this case also the frequency is involved in the decomposition in a nontrivial way. The resulting asymptotics at $\mathscr I^-$ are then given by
\begin{equation}\label{}
	p_{\omega l m} \sim \int_0^{\infty} d\omega' \sum_{l'm'} c^{(+)}_{\omega \omega' l l' m m'}\left[\frac{t_{\omega'l'}}{\sqrt{2\pi\omega'}\, r}\, e^{i\omega' u(v)} \,Y_{l'm'}(\xi) \Theta(v_0-v)\right]
	\qquad (\text{at }\mathscr I^-)\,,
\end{equation}
or equivalently
\begin{equation}\label{}
	p_{\omega l m} \sim \frac{1}{\sqrt{2\pi\omega}\, r} \int d\Omega(\xi') Y_{lm}(\xi'_+)  R_+(\xi')  t_{\omega R_+(\xi')}(\xi,\xi') \left(\frac{v_0-v}{C}\right)^{-\frac{i}{\kappa}\omega R_+(\xi')} \quad (\text{at }\mathscr I^-)\,,
\end{equation}
where we adapted the definition \eqref{smearedt} to the current notation.
The Bogolyubov coefficients then take the form
\begin{equation}\label{}
	\begin{aligned}
		\alpha_{\omega \omega' l l' m m'}
		=\ &
		\frac{1}{2\pi n}\sqrt{\frac{\omega'}{\omega}}
		\int d\Omega(\xi) Y_{lm}(\xi_+) R_+(\xi) 
		\int d\Omega(\xi') Y_{l'm'}^\ast(\xi'_+) R_-^2(\xi')\\
		& \times t_{\omega R_+(\xi)} (\xi , \xi')
		C^{\frac{i}{\kappa}\omega R_+(\xi)} e^{-i\omega' R_-(\xi')v_0} \frac{\Gamma\left(1-\frac{i}{\kappa}\omega R_+(\xi)\right)}{(\epsilon-i\omega' R_-(\xi'))^{1-\frac{i\omega}{\kappa}R_+(\xi)}}\,.
	\end{aligned}
\end{equation}
and
\begin{equation}\label{key}
	\begin{aligned}
		\beta_{\omega \omega' l l' m m'}
		=\ &
		\frac{1}{2\pi n}\sqrt{\frac{\omega'}{\omega}}
		\int d\Omega(\xi) Y_{lm}(\xi_+) R_+(\xi) 
		\int d\Omega(\xi') Y_{l'm'}^\ast(\xi'_+) R_-^2(\xi')\\
		& \times t_{\omega R_+(\xi)} (\xi , \xi')
		C^{\frac{i}{\kappa}\omega R_+(\xi)} e^{i\omega' R_-(\xi')v_0} \frac{\Gamma\left(1-\frac{i}{\kappa}\omega R_+(\xi)\right)}{(\epsilon+i\omega' R_-(\xi'))^{1-\frac{i\omega}{\kappa}R_+(\xi)}}\,.
	\end{aligned}
\end{equation}
Putting everything together we finally arrive at the two-point function
\begin{equation} 
	\begin{split}
	\left\langle \boldsymbol b^\dagger_{\omega l m} \,\boldsymbol b_{\omega'l'm'}  \right\rangle  & = \iint \frac{ d\Omega(\xi_+)  d\Omega(\xi'_+) }{n\sqrt{R_+(\xi) R_+(\xi')}}\,  Y_{lm}(\xi_+)\,\frac{\Gamma_{\omega R_+(\xi)}(\xi, \xi')}{e^{\frac{2\pi \omega R_+(\xi)}{\kappa}}-1}\,Y_{l'm'}^\ast(\xi'_+) \\ & \times \, \delta(\omega R_+(\xi)-\omega' R_+(\xi')) \,, \label{mainSR}
	\end{split}
\end{equation}
which again displays non-trivial correlations between different $l,l'$ and $m,m'$ modes. Note that the previous equation could have been obtained directly from  
the corresponding expression \eqref{bbSR qbasis} in the $\omega,\hat q$ basis by using the relation \eqref{changeofbasis} for the creation and annihilation operators. 
Moreover, restricting to diagonal elements and summing over angular labels again leads to the spectral emission rate \eqref{dNtotSR}.

Just like for supertranslations, \eqref{mainSR} greatly simplifies when approximating $|t_{\omega l}|^2\simeq |t_{\omega}|^2$ (whose validity was discussed in Section \ref{sec: supertranslations}). In that case, it becomes
\begin{equation}\label{}
	\left\langle \boldsymbol b^\dagger_{\omega l m} \,\boldsymbol b_{\omega'l'm'}  \right\rangle \simeq   \int d\Omega(\xi_+) \, Y_{lm}(\xi_+)\, \frac{|t_{\omega R_+(\xi)}|^2}{e^{\frac{2\pi \omega R_+(\xi)}{\kappa}}-1}\,Y_{l'm'}^\ast(\xi_+)\,\delta(\omega -\omega') \,,
\end{equation}
which further simplifies in the pure-absorption scenario, $|t_{\omega l}|^2\simeq 1$, to
\begin{equation}\label{SRpureabs}
	\left\langle \boldsymbol b^\dagger_{\omega l m} \,\boldsymbol b_{\omega'l'm'}  \right\rangle \simeq   \int d\Omega(\xi_+) \, \frac{ Y_{lm}(\xi_+)Y_{l'm'}^\ast(\xi_+)}{e^{\frac{2\pi \omega R_+(\xi)}{\kappa}}-1}\,\delta(\omega -\omega') \,.
\end{equation}

\subsection{High-energy behavior of the correlators in $\omega,l,m$ basis}
Let us now look more carefully at the behavior of the correlators in the $\omega,l,m$ basis. In order to make the  results more transparent let us focus on the pure-absorption scenario \eqref{SRpureabs} and focus on a specific family of boosts and superrotations defined by 
\begin{equation}\label{SR+1}
(z,\bar{z}) \rightarrow \left(G_+(z),\tilde {G}_+(\bar{z})\right)=\left(c z^n,c \bar{z}^n\right) \,, \qquad c\in\mathbb R
\end{equation}
in terms of stereographic coordinates,
which translates to
\begin{equation}\label{SR+2}
\phi_+=n\phi\,,\qquad \theta_+=2\arctan\left[
c \left(\tan\frac{\theta}{2}\right)^{\!\!n}\,
\right]
\end{equation}
in standard spherical coordinates. Note that, if the original azimuthal angle $\phi$ varies from $0$ to $2\pi$, the new coordinate $\phi_+$ can wind up to $n$ times around the sphere, producing a multiple cover. It is to compensate for this fact that we introduced factors of $n$ where needed in the above equations.
In Appendix \ref{app:conformal}  we derive in more detail the form of $R_+(\xi)$ for local conformal transformations. Here we reproduce the final result
 \begin{equation} \label{R general form}
R_+(z,\bar z) =c\, n \frac{|z|^{n-1}(1+|z|^2)}{1+c^2|z|^{2n}} =  
 \frac{n}{2}\, c^{\frac{1}{n}}(1+x)  \left(\frac{1-x}{1+x}\right)^{\frac{n-1}{2n}} 
 \left[
 1+\frac{1}{c^{\frac{2}{n}}}\left(\frac{1-x}{1+x}\right)^{\!\!\frac{1}{n}} \,
 \right] \, ,
 \end{equation}
 where $x=\cos(\theta_+)$. 
 The case $n=1$ corresponds to boosts in the directions $\theta=0,\pi$ with velocity $v$ (positive or negative) and
 \begin{equation}\label{rapidityandgamma}
 	c=\sqrt{\frac{1+v}{1-v}}\,\qquad \gamma=\frac{1}{\sqrt{1-v^2}}\,.
 \end{equation}
The associated conformal factor takes the simple form
 \begin{equation} \label{RB}
 \text{\textit{{Boosts}}:} \qquad    R_+ = c_+   +   x c_-\,,\qquad c_\pm \equiv \frac{1}{2}\left(c\pm \frac{1}{c}\right)\,. 
 \end{equation}
Note that $c=e^\psi$ in terms of the rapidity $\psi$, so that $c_+=\cosh\psi$ and $c_-=\sinh\psi$. 
For any finite, nonzero $c$, this conformal factor is strictly positive for $|x|\le 1$ because its only root is at $x=\frac{1+c^2}{1-c^2}$, which never satisfies $|x|\le1$.
This root however approaches one of the poles in the case of ultrarelativistic boosts where $c\to\infty$ or $c\to0$.

On the other hand,
for $n=2,3,\ldots$ we have superrotations whose conformal factors  always asymptote to zero near the poles of the sphere, according to 
\begin{equation}\label{asymptRtilde}
\begin{aligned}
\text{\textit{{Superrotations}}:} \qquad    R_+ (x) \,&\sim\,  {n} \begin{cases}
c^{\frac{1}{n}}\left(\frac{1-x}{2}\right)^{\frac{n-1}{2n}}\qquad  &(x\to 1^-)\\
c^{-\frac{1}{n}}\left(\frac{1+x}{2}\right)^{\frac{n-1}{2n}}\qquad  &(x\to -1^+)\, .
\end{cases} 
\end{aligned}
\end{equation}
Values of $c$ different from unity give thus rise to an asymmetry between northern  and southern hemisphere.

The fact that there is no dependence on $\phi$ in the conformal factors under considerations allows us to further simplify eq. \eqref{SRpureabs} to
\begin{equation} \label{bdaggerb n}
\left< \boldsymbol b^\dagger_{\omega l m} \boldsymbol b_{\omega'l'm'}  \right> \simeq   \delta\left(\omega-\omega'\right) \delta_{m m'}  \,s_{lm} s_{l'm} \, n_{ll'm}(\omega)\,,
\end{equation}
where $s_{lm}$ is the normalization coefficient given in eq. \eqref{spherical harmonic expansion} and we isolated the following integral
\begin{equation}\label{transllm}
n_{ll'm}  (\omega)=\int_{-1}^{+1}
\frac{
	P_l^m(x)
	P_{l'}^m(x)}{
	e^{\tilde \omega  R_+(x)}-1}\,dx\,,\qquad \tilde \omega\equiv \frac{2\pi \omega}{\kappa}\,,
\end{equation}
with $P_l^m(x)$ the associated Legendre polynomials (see Appendix \ref{app: spherical harmonics}). Note that in eq. \eqref{R general form} $R_+$ is expressed in terms of the new coordinate system $x=\cos(\theta_+)$ and therefore in eq. \eqref{transllm} all functions are evaluated at the same arguments.
The integral in eq. \eqref{transllm} is in general hard to solve for $R_+$ given by eq. \eqref{R general form}. Therefore, we study its high-energy limit, which is also the regime in which the pure-absorption approximation is better justified.

\subsubsection{Mild boosts}
\label{ssec: HELboosts}

Let us start by considering the high-energy limit $\tilde \omega \gg 1$ of the integral in eq.~\eqref{transllm} in the case $n=1$ associated to standard boosts. We will assume the boost factor to be sufficiently mild, so that $\tilde \omega \gg c, 1/c$, postponing the analysis of ultrarelativistic boosts to Section \ref{sec:URBoost}. According to eq. \eqref{RB}, $R_+$ is strictly positive and hence we can perform a large-frequency expansion neglecting the $-1$ in the denominator of  \eqref{transllm},
\begin{equation}\label{}
n_{ll'm} (\omega)
\simeq
\int_{-1}^{+1}
P_l^m(x)
P_{l'}^m(x)
e^{-\tilde \omega  R_+(x)}\,dx
=
e^{-c_+\tilde\omega}\int_{-1}^{+1}
P_l^m(x)
P_{l'}^m(x)
e^{-c_-\tilde \omega x}\,dx\, .
\end{equation}
Integrating by parts, this gives
\begin{equation}\label{tIboost}
n_{ll'm}  (\omega)
\simeq e^{-c_+\tilde \omega}\sum_{k} \frac{b^-_ke^{c_-\tilde \omega}-b^+_ke^{-c_-\tilde \omega}}{(c_-\tilde \omega)^{k+1}}
\end{equation}
provided the boost is nontrivial, $c\neq 1$.
Here $b_{k}^\pm$ denote coefficients given by the derivatives
\begin{equation}\label{}
b_k^\pm = \frac{d^k}{dx^k}(P_l^{m}P_{l'}^m)\Big|_{x=\pm1} \qquad  (k=0,1,2,\ldots)
\end{equation} 
which thus depend on $l$, $l'$ and $m$. 
Looking at the asymptotics of the Legendre polynomials near the poles, eqs. \eqref{Legendre polynomial near the poles} and \eqref{reflection law}, we see that the first nontrivial $b_k^\pm$ actually occurs for $k=|m|$ and is given by
\begin{equation}\label{bcllm}
b_{|m|}^\pm=|m|!\  c^\pm_{ll'm}\,,\qquad
c_{ll'm}^\pm\equiv\lim_{x\to\pm 1}\frac{P_l^m(x)P_{l'}^m(x)}{(1-|x|)^{|m|}}\,.
\end{equation}
For $c>1$, which corresponds to a boost in the  $\theta=0$ direction, we can neglect the contribution proportional to $e^{-c_-\tilde\omega}$ in \eqref{tIboost} up to an exponentially small error so that
\begin{equation}\label{eq:Boosts Particle number c>1}
n_{ll'm} (\omega)\simeq e^{-\frac{\tilde\omega}{c}}\sum_{k\ge|m|}\frac{b_k^-}{(c_-\tilde\omega)^{k+1}}=\frac{e^{-\frac{\tilde \omega}{c}}}{(c_- \tilde\omega)^{|m|+1}}
|m|!\ c_{ll'm}^{-}
\left(
1+\mathcal O(\tilde \omega^{-1})
\right)\,.
\end{equation}
Similarly, for $0<c<1$,
\begin{equation}\label{eq:Boosts Particle number c<1}
n_{ll'm}  (\omega)\simeq -e^{-c\,\tilde\omega}\sum_{k\ge|m|}\frac{b_k^+}{(c_-\tilde\omega)^{k+1}}=-\frac{e^{-c\,\tilde \omega}}{(c_- \tilde\omega)^{|m|+1}}
|m|!\ c_{ll'm}^{+}
\left(
1+\mathcal O(\tilde \omega^{-1})
\right)\,.
\end{equation}
In both cases we see that the boosted spectrum retains the standard Boltzmann suppression of the original spectrum but now with a frequency rescaled by the boost factor. There are also nonzero off-diagonal entries in $l,l'$ space which are however exponentially suppressed. 

\begin{figure}
	\centering
	\includegraphics[width=0.47\linewidth]{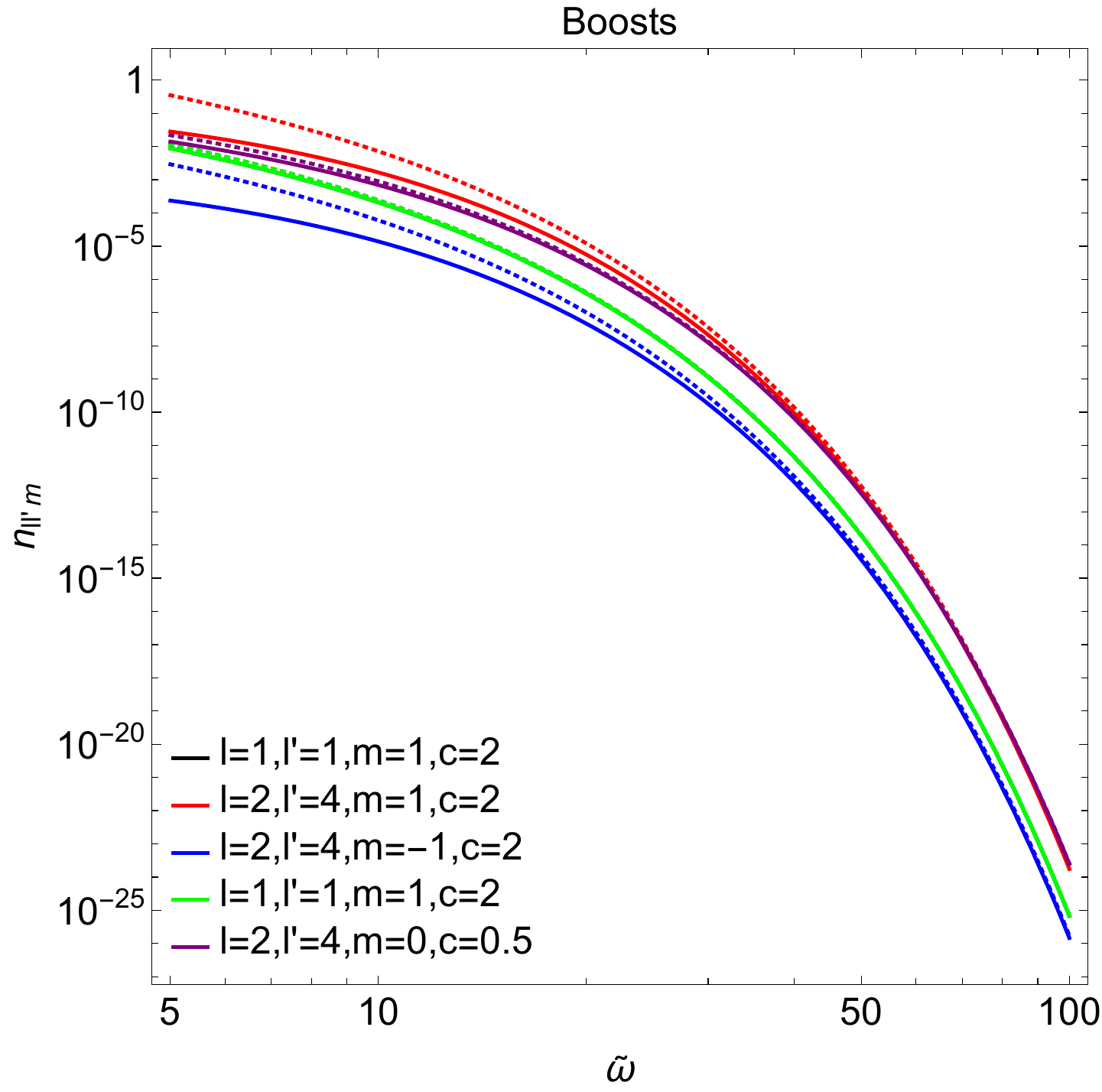} \,
	\includegraphics[width=0.47\linewidth]{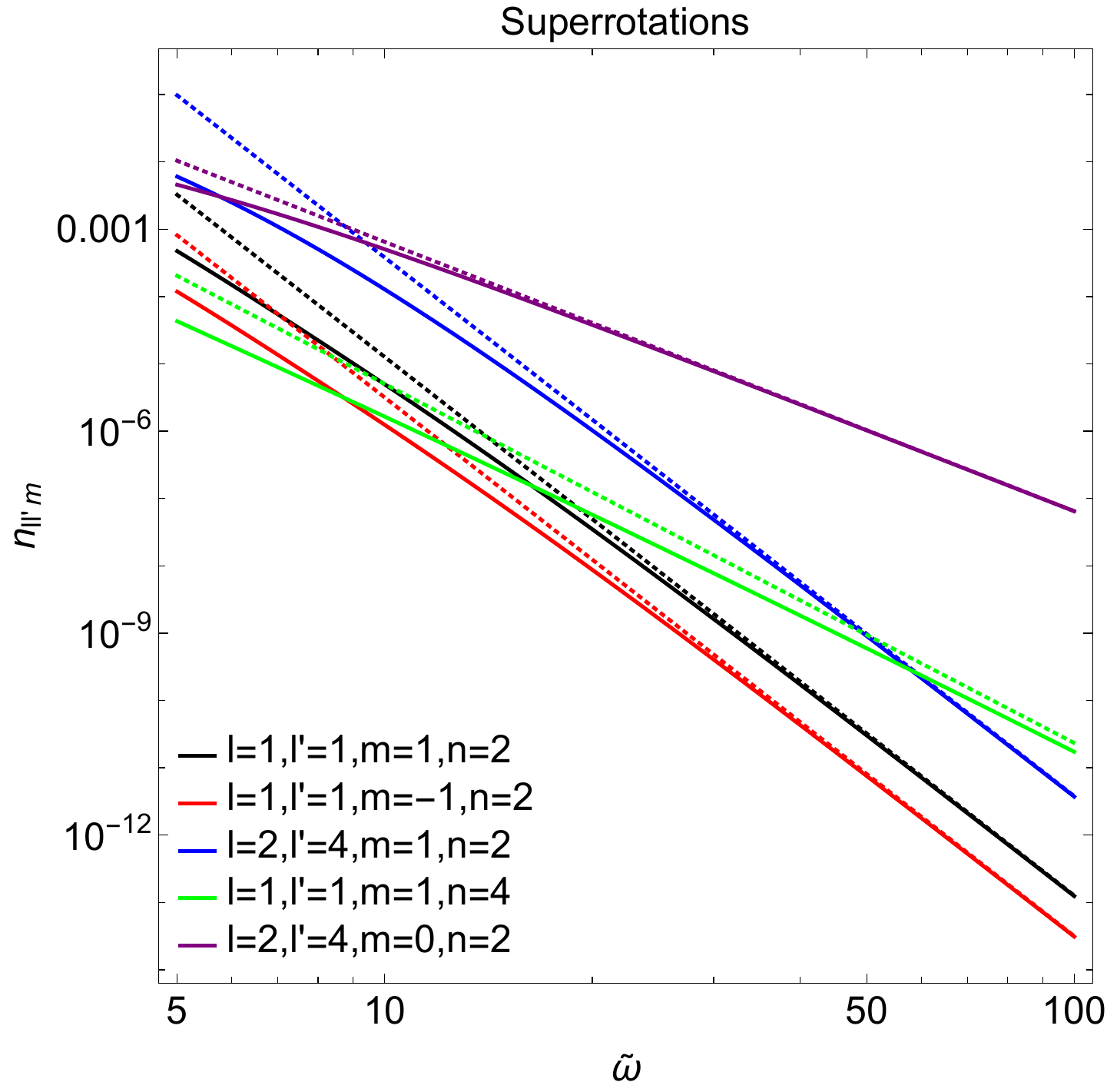}
	\caption{The matrix elements $n_{ll'm}$ for boosts (left) and superrotations (right) for different choices of $l,l',m$ as a function of $\tilde \omega = 2\pi \omega / \kappa$. Solid lines are the numerical integrals and dashed lines are the analytical approximations using, for boosts, the leading terms in eqs. \eqref{eq:Boosts Particle number c>1} and \eqref{eq:Boosts Particle number c<1}, and for superrotations, eq. \eqref{t final SR}.}
	\label{fig:bbdaggersr}
\end{figure}

In the left plot of Figure \ref{fig:bbdaggersr} we compare numerical results with the first terms in the analytical approximations in eqs. \eqref{eq:Boosts Particle number c>1} and \eqref{eq:Boosts Particle number c<1}. The approximation is accurate up to relatively small frequencies, say $\tilde\omega\simeq 10-20$ (equivalently $\omega/\kappa \simeq 2-4$). In particular let us stress again the main feature, the exponential suppression of the spectrum for large frequencies.

\subsubsection{Superrotations}
\label{ssec: HELsuperrotations}

\begin{figure}
	\centering
	\begin{tikzpicture}
	\draw[help lines,->] (-5,0) -- (5,0) coordinate (xaxis);
	\draw[help lines,->] (0,0) -- (0,7) coordinate (yaxis);
	\path [fill=red!50!white] (-4.22,0)--(-4.22,1) to [out=-110,in=80] (-4.5,0);
	\path [fill=red!50!white] (4.22,0)--(4.22,1) to [out=-70,in=100] (4.5,0);
	\draw (-4.5,0) to [out=80,in=180] (0,4);
	\draw (0,4) to [out=0,in=100] (4.5,0);
	\draw [dashed] (-5,1) -- (5,1);
	\draw [dashed] (-5,.1) -- (5,.1);
	\draw [dashed] (-5,6) -- (5,6);
	\draw[fill] (4.22,0) circle [radius=0.03];
	\draw[fill] (-4.22,0) circle [radius=0.03];
	\draw[fill] (4.5,0) circle [radius=0.03];
	\draw[fill] (-4.5,0) circle [radius=0.03];
	\node at(-3.9,0)[below]{$-a$};
	\node at(4.2,0)[below]{$a$};
	\node at(-4.7,0)[below]{$-1$};
	\node at(4.6,0)[below]{$1$};
	\node at(5,0)[below]{$x$};
	\node at(0,4.3)[left]{$R_+(x)$};
	\node at(-5,6)[left]{$1$};
	\node at(-5,1)[left]{$R_+(a)$};
	\node at(-5,.1)[left]{${\tilde\omega}^{-1}$};
	\end{tikzpicture}
	\caption{A schematic representation of $R_+(x)$ for superrotations with $c=1$. The shaded area highlights Region II, while the rest of the $[-1,1]$ interval corresponds to Region I. The dashed lines illustrate the separation of scales considered in eq.~\eqref{ascale}.}
	\label{fig:Rdisegno}
\end{figure}
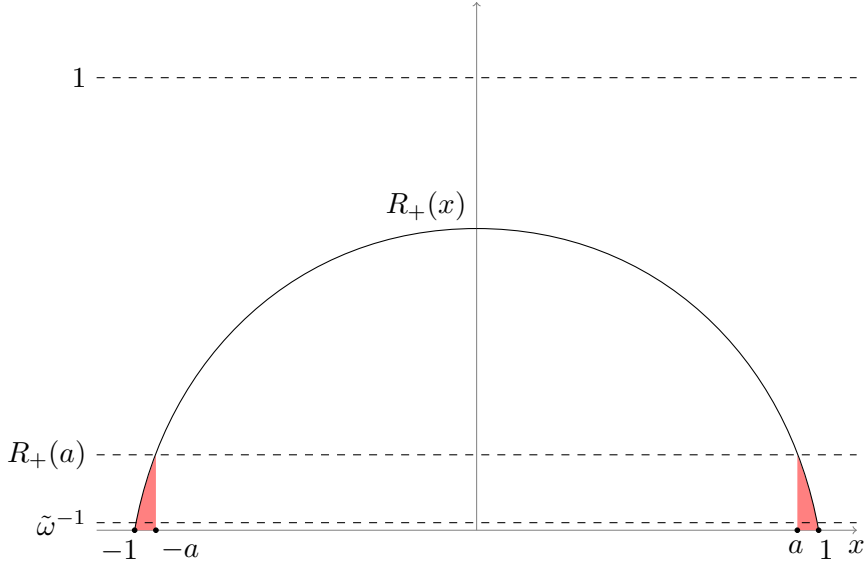

For $n=2,3,4,\ldots$ , the conformal factor $ R_+(x)$ vanishes near the poles according to  eq. \eqref{asymptRtilde}. Therefore we need to distinguish two regions as $\omega\to\infty$ depending on whether $\tilde \omega R_+(x)$ is large or not. To do this consider a value $a$ such that 
\begin{equation}\label{ascale}
\frac{1}{\tilde \omega}\ll  R_+(a) \ll 1\,.
\end{equation} 
In practice, $a$ should be sufficiently close to 1 so that the condition $a<|x|<1$ identifies points near the north and south pole, where $R_+(x)$ is indeed small, as depicted in Figure \ref{fig:Rdisegno}.
Then, splitting the integration interval accordingly,
\begin{equation}\label{}
\int_{-1}^{1}dx=\underbrace{\int_{|x|<a}dx}_\text{Region I}+\underbrace{\int_{a<|x|<1}dx}_\text{Region II}
\end{equation}
defines the two regions of interest. Note that Region II is a genuinely new feature of superrotations compared to standard boosts. As we will see in the next section, however, a similar behavior is also shared by ultrarelativistic boosts.

In Region I, we are allowed to regard $\tilde \omega  R_+(x)$ as a large quantity, expanding the denominator in eq.~\eqref{transllm},
\begin{equation}
n_{ll'm} ^{(\mathrm{I})} (\omega)
=\int_{|x|<a}
\frac{
	P_l^m(x)
	P_{l'}^m(x)}{
	e^{\tilde \omega  R_+(x)}-1}\,dx 
\simeq
\int_{|x|<a}
P_l^m(x)
P_{l'}^m(x)
e^{-\tilde \omega  R_+(x)}\,dx\,.
\end{equation}
This quantity has the following upper bound,
\begin{equation}\label{exp sup}
|n_{ll'm} ^{(\mathrm{I})} (\omega)| < \max_{x\in[-1,1]}
\left| 
P_l^m(x)
P_{l'}^m(x)
\right|\, e^{-\tilde \omega R_+(a)}\,,
\end{equation} 
which displays an exponential suppression for large $\tilde \omega$ thanks to eq.~\eqref{ascale}.

In Region II, we obtain
\begin{equation}\label{canccpcm}
n_{ll'm} ^{(\mathrm{II})} (\omega)
=\int_{a<|x|<1}
\frac{
	P_l^m(x)
	P_{l'}^m(x)}{
	e^{\tilde \omega  R_+(x)}-1}\,dx 
\simeq
c^-_{ll'm}
\int_{-1}^{-a}
\frac{(1+x)^{|m|}}{
	e^{\tilde \omega  R_+(x)}-1
}\,dx
+
c^+_{ll'm}
\int_{a}^{1}
\frac{(1-x)^{|m|}}{
	e^{\tilde \omega  R_+(x)}-1
}\,dx\,,
\end{equation}
where the coefficients $c^{\pm}_{ll'm}$ are defined in eq. \eqref{cpm}.
Since $a$ is very close to 1, we have substituted the numerator with its asymptotic value near the poles, while the denominator is unexpanded. Furthermore, approximating $ R_+(x)$ as in \eqref{asymptRtilde} we find 
\begin{equation}\label{tIISRnum}
n_{ll'm} ^{(\mathrm{II})} (\omega)\simeq
c^-_{ll'm} \int_0^{\infty}
\frac{y^{|m|}}{
	\exp\left[n\tilde\omega c^{\frac{1}{n}}\left(\frac{y}{2}\right)^{\frac{n-1}{2n}}\right]-1
}\,dy
+
c^+_{ll'm} \int_0^{\infty}
\frac{y^{|m|}}{
	\exp \left[n\tilde\omega c^{-\frac{1}{n}}\left(\frac{y}{2}\right)^{\frac{n-1}{2n}}\right]-1
}\,dy\,,
\end{equation}
where we have extended the integration range to infinity, which is justified up to an exponentially small error. Hence, after evaluating the leftover integrals using
\begin{equation}\label{GammaZeta}
\int_0^\infty\frac{x^{\alpha-1}}{e^x-1}\,dx=\Gamma(\alpha)\zeta(\alpha)\,,
\end{equation}
where $\zeta$ is the Riemann zeta function, we obtain
\begin{equation}\label{t final SR}
n_{ll'm} ^{(\mathrm{II})} (\omega)
\simeq
(c^-_{ll'm}c^r+c^+_{ll'm}c^{-r})\frac{2^{|m|+1}}{(n\tilde\omega)^{nr}}
\frac{2n}{n-1}
\Gamma(nr)\zeta(nr)\,,\qquad r\equiv\frac{2(|m|+1)}{n-1}\, .
\end{equation}
Thus, by comparing with eq. \eqref{exp sup} we see that Region II dominates the integral.
For $c=1$ there is an exact cancellation whenever $l+l'$ is odd because the integrand is anti-symmetric between the north and south pole due to eq. \eqref{reflection law} or, at first order, due to the relation
\begin{equation}\label{}
c_{ll'm}^-= c_{ll'm}^+ (-1)^{l+l'}\,.
\end{equation}

On the right plot of Figure \ref{fig:bbdaggersr} we show that eq. \eqref{t final SR} is in good agreement with the numerical results for  $\tilde{\omega}\gtrsim 10-20$. In particular, we can confirm the power-law behavior for large frequencies. 

The power-law behavior $1/\omega^{nr}$ arising from Region II and controlled by the superrotation parameter $n$ is the main novel feature of superrotations. It reflects the fact that the modes close to the poles are redshifted in a particular way dictated by superrotations. The power-law behavior is distinctive and differs from the high-energy limit of mild boosts, which display a Boltzmann suppression at large frequencies, and from the case of ultrarelativistic boosts which, as we will see next is suppressed by $\omega^{-1}\log\omega$ for $m=0$ and $1/\omega^{|m|+1}$ for $|m|>0$. It is also different from the low-energy limit $\omega/\kappa \ll 1$, where the two-point function is generically suppressed as $1/\omega$, and so it is distinguishable from a large redshift of the standard behavior.

The singular nature of the superrotations, whose conformal factor vanishes at the poles, may however cast some doubts concerning its actual physical realizability. To address this point, let us observe that the integral is finite and that the integrand actually peaks at an intermediate region where $\omega R_+ \simeq \kappa/(2\pi)$.  Therefore, by introducing a cutoff sufficiently near to the poles, the onset of the exponential decay can be delayed to higher and higher frequencies, thus suggesting the existence of a regime in which the power-law decay is indeed physical. In Section \ref{sec: power spectrum} we will study which type of superrotations can lead to physically sensible power spectrum.

\subsubsection{Ultrarelativistic boosts}
\label{sec:URBoost}

Let us reconsider here the integrals associated to boosts and discuss the ultrarelativistic limit. 
Note that $R_+(x)$ in eq. \eqref{RB} is bounded from below by $R_+(x)> \operatorname{min}[c,1/c] $. Therefore, if the boost factor $c$ is either very large or very small, then, even in the high-energy limit $\tilde\omega \gg 1$, there will be regions in the angular integral where $\tilde{\omega} R_+ \ll 1$. 

Let us focus for definiteness on the case $1\ll \tilde{\omega} \ll c$. 
The case with $1\ll \tilde{\omega} \ll 1/c $ is equivalent up to interchanging the north with the south pole.
The splitting between Region I and Region II can be defined, for instance, by the value $x=-1+1/c$. In Region I, namely $-1+1/c<x<1$, we are allowed to regard $\tilde \omega  R_+(x)$ as a large quantity. Then, after expanding the denominator in eq.~\eqref{transllm} we get the usual exponential suppression thanks to eq.~\eqref{ascale}.
In Region II, namely $-1<x<-1+1/c$, we can retrace the steps we followed in the case of superrotations to obtain, up to an exponentially small error,
\begin{equation}\label{}
	n_{ll'm} ^{(\mathrm{II})} (\omega)
	=\int_{-1}^{-1+\frac{1}{c}}
	\frac{
		P_l^m(x)
		P_{l'}^m(x)}{
		e^{\tilde \omega  R_+(x)}-1}\,dx 
	\simeq
c^-_{ll'm}\left(\frac{2}{\tilde \omega c}\right)^{|m|+1}
\int_{0}^{+\infty}
\frac{y^{|m|}}{
	e^{\frac{\tilde\omega}{c}+ y}-1
}\,dy\,,
\end{equation}
where the coefficients $c^{\pm}_{ll'm}$ are defined in eq. \eqref{cpm}. 
For $|m|=1,2,3,\ldots$ we can neglect the $\tilde \omega/c$ appearing in the denominator, finding to leading order
\begin{equation}
	n_{ll'm} ^{(\mathrm{II})} (\omega)
\simeq
c^-_{ll'm}\left(\frac{2}{\tilde \omega c}\right)^{|m|+1}
\Gamma\left(|m|+1\right)\zeta\left(|m|+1\right)
\end{equation}
after using \eqref{GammaZeta}.
The integral is instead elementary for $m=0$,
\begin{equation}
\int_0^\infty
\frac{dy}{
	e^{\frac{\tilde \omega}{c}+ y}-1
} = - \log\left(1-e^{-\frac{\tilde \omega}{c}}\right)
\end{equation}
and thus for the regime of interest we find
\begin{equation}
n_{ll'0} ^{(\mathrm{II})} (\omega)
\simeq
c^-_{ll'0}\, \frac{2}{\tilde \omega c}\,\log \left(\frac{c}{\tilde \omega}\right)\,.
\end{equation}
As anticipated, a high-speed boost modifies the spectrum by inducing a power-law decay at high frequencies, $1/\omega^{|m|+1}$ for $m\neq 0$, or a polyhomogenous behavior $\omega^{-1}\log\omega$ for $m=0$.

\section{Power Spectrum and Particle Emission Rate}
\label{sec: power spectrum}

Finally let us look at the power spectrum and emission rate and study how these observables are affected by superrotations. As we showed in Section~\ref{sec: supertranslations}, the case of supertranslations only differs from the standard case for off-diagonal entries in the two-point function and so its possible signatures are not captured by these observables.
 
The power emitted by the black hole per unit of frequency for each $l$ and $m$ mode is related to the spectral rate of emitted particles by 
\begin{equation}\label{powerspectrum}
	dP_{\,lm}(\omega) =  \omega\, dN_{lm}(\omega)\,.
\end{equation}
Let us start by revisiting the power spectrum in the standard Hawking case where the spectral emission rate is given by eq. \eqref{Hawking rate}, so that
\begin{equation} \label{eq: power spectrum per unit  frequency}
	dP(\omega) = \sum_{lm} dP_{lm}(\omega) = \frac{1}{2\pi} \frac{g(\omega)}{e^{\frac{2\pi\omega}{\kappa}}-1}\,\omega\, d\omega \,,
\end{equation}
where $g(\omega)$ is given in eq.~\eqref{dos}.
In the high-energy regime, $2\pi\omega /\kappa \gg 1$, we expect the geometric-optics limit to be applicable, so that the transmission probabilities $|t_{\omega l}|^2$ can be approximated by perfect absorption for $\omega \gtrsim l/R_a$ and perfect reflection for $\omega \lesssim l/R_a$, i.e. 
\begin{equation} \label{transmission coefficients}
	|t_{\omega l}|^2 \simeq \Theta(\omega R_a-l)\,,
\end{equation}
where $A=4 \pi R_a^2$ is the effective area of the emitting body, with $R_a=3\sqrt{3} r_s/2$ \cite{Page:1976df,Giddings:2015uzr}. Therefore, in this regime, the density of states \eqref{dos} can be approximated as
\begin{equation}\label{eq: density of states}
	g(\omega)\approx (\omega R_a)^2\,.
\end{equation}
As we explain in Appendix~\ref{app: Transmission coefficients}, for a massless minimally coupled scalar field the density of states for any frequency turns out not to differ more than a factor of $26/11$ from this result and therefore it is a good approximation to use eq. \eqref{eq: density of states} for all frequencies. This way we get
\begin{equation}\label{Hpowerspectrum}
	dP(\omega)\simeq \frac{1 }{2\pi} 
	\frac{(\omega R_a)^2}{e^{\frac{2\pi\omega}{\kappa}}-1}\,\omega\, d\omega \,  .  
\end{equation}
An estimate for the total emitted power can be then obtained by integrating the previous equation over all frequencies which yields the approximate expression
\begin{equation}
	P_H \simeq \frac{1}{2}\, \sigma_s \, A \, T^4 \,,\qquad \sigma_s = \frac{\pi^2}{60}\,.
\end{equation}
The black hole emission spectrum thus satisfies the Stefan-Boltzman law for a radiating black body  with a temperature $T=\kappa/(2\pi)$ except for an overall factor of $1/2$, which is due to the fact that the scalar field only carries one degree of freedom, in contrast with the electromagnetic field.
Proceeding along the same lines for the particle emission rate affords instead
\begin{equation}\label{}
	dN(\omega)\simeq \frac{(\omega R_a)^2 }{2\pi} 
	\frac{1}{e^{\frac{2\pi\omega}{\kappa}}-1}\, d\omega 
\end{equation}
and
\begin{equation}\label{}
	N_H = \frac{\zeta(3)}{4\pi^2} A T^3\,.
\end{equation}

The case of boosts and superrotations is more interesting, due to the nontrivial dependence on the conformal factor. This time is more convenient to work with the basis where the states are labelled by a direction $\hat{q}$ rather than by the quantum numbers $l,m$. 
Starting from \eqref{dNSR} and integrating over frequencies gives the emission rate per solid angle
\begin{align}\label{}
	dN (\hat q) &= \frac{d\Omega(\hat q)}{8\pi^2}\,\int \frac{g(\omega_+)}{e^{\frac{2\pi\omega_+}{\kappa}}-1} \frac{1}{R_+^2(\hat Q_+)}  \,d\omega \, ,
\end{align}
whereas if we multiply by $\omega$ before integrating yields the power emitted per solid angle
\begin{align}\label{}
	dP(\hat q) &=     \frac{d\Omega(\hat q)}{8\pi^2}\,\int  \frac{g(\omega_+)}{e^{\frac{2\pi\omega_+}{\kappa}}-1} \frac{1}{R_+^2(\hat Q_+)}\,  \omega \,d\omega  \, . \label{PPSR}
\end{align}
Then, approximating the density of states by \eqref{eq: density of states} and integrating over the angles yields the total power and emission rate
\begin{equation}\label{}
	P \simeq \frac{1}{4\pi} \oint \frac{d\Omega(\hat k_+)}{R_+^4(\hat k)}\, P_H\,,\qquad
	N \simeq \frac{1}{4\pi} \oint \frac{d\Omega(\hat k_+)}{R_+^3(\hat k)}\, N_H\,.
\end{equation}

For boosts with conformal factor dictated by \eqref{RB}, the total power and particle rate detected by the boosted observer take the form 
\begin{equation}
P_B \simeq \frac{1}{3}(v^2+3)\, \gamma^2 P_H\,,\qquad
N_B \simeq \gamma N_H\,,
\end{equation}
where $v$ is the velocity and $\gamma$ is the relativistic factor \eqref{rapidityandgamma}. These results are in complete analogy with the case of a boosted blackbody spectrum \cite{Ford:2013koa}.
Accordingly, for an ultrarelativistic boost $c\gg 1$, these quantities grow as
\begin{equation}
P _B\,\sim\, \frac{c^2}{3} P_H\,,\qquad
N_B \,\sim\, \frac{c^2}{4} N_H\,.
\end{equation}
For superrotations $(z,\bar z)\mapsto c \,(z^n, \bar z^n)$, the conformal factor vanishes at the poles following \eqref{asymptRtilde}.  Therefore, by placing a small cutoff $\Delta$ surrounding the poles, we can discuss the convergence of the corresponding power and emitted particle rate by inspecting
\begin{equation}\label{}
P_{SR}
\,\propto \,\int_\Delta^1\frac{dx}{x^{\frac{2(n-1)}{n}}}\,,\qquad
N_{SR} \,\propto \,\int_\Delta^1\frac{dx}{x^{\frac{3(n-1)}{2n}}}\,.
\end{equation}
The power spectrum diverges logarithmically for $n=2$ and as $\Delta^{-\frac{n-2}{n}}$ for $n>2$. The emitted particle rate $N$ is instead finite for $n=2$,
explicitly given by
\begin{equation}
N_{SR}=\left[\frac{3}{8} \sqrt{c} \left(c^2+1\right) \cot ^{-1}\left(\sqrt{c}\right)+\frac{c \left(-3 c^3-5
	c^2+5 c+3\right)}{8 (c+1)^2}\right]N_H\,,
\end{equation} 
while it diverges logarithmically for $n=3$ and as $\Delta^{-\frac{n-3}{2n}}$ for $n>3$.
 
However, it might be physically more transparent to rephrase these divergences in terms of ultraviolet sensitivity. Then, we return to eq. \eqref{dNSR}, multiply by the frequency and perform first the integral over the sphere. By approximating the density of states as $g(\omega R_+(\hat Q)) \simeq (\omega  R_+(\hat Q) R_a)^2$ for all frequencies, as we justified above, we can then follow the steps in Section \ref{ssec: HELsuperrotations} where it was noted that for $\omega \gg \kappa/(2\pi)$ the integral is dominated by the regions near the poles where $\omega R_+ \sim \kappa/(2\pi)$. Using eq. \eqref{asymptRtilde} we thus find an expression analogous to eq. \eqref{t final SR} namely
\begin{equation} 
	P_{SR}(\omega) 
	\simeq \frac{	n \tilde{ r}}{2 \pi^2 n^{n\tilde{r}}}
	(c^{\tilde{r}}+c^{-\tilde{r}})
	\Gamma(n\tilde{r})\zeta(n\tilde{r}) R_a^2 \int_{T}^\Lambda  \frac{\omega^3}{\tilde\omega^{n\tilde{r}}} \, d\omega \,,\qquad \tilde{r} \equiv\frac{2}{n-1}\, .
\end{equation}
where we have inserted an ultraviolet cutoff at a frequency much larger than the Hawking temperature, $\Lambda \gg T=\kappa/(2\pi)$, and an infrared cutoff at $T$, where the high-energy approximation used in performed the angular integral becomes unreliable. After integration we find the leading terms to be
\begin{align} \label{power spectrum superrotations}
	P_{SR} \simeq  
	\begin{cases}
	\dfrac{\pi}{480} \left(c^2+c^{-2}\right)    A \,T^4 \log\left(\dfrac{\Lambda}{T}  \right) \,  \qquad & (n=2)
	\\ 
	\\
	\dfrac{n \tilde{ r} }{8\pi^3 }\zeta (n \tilde{r}) \Gamma ( n \tilde{r}) \left(c^{\tilde{r}}+c^{-\tilde{r}}\right)  A \left(\dfrac{T}{n \Lambda}\right)^{n \tilde{r}} \dfrac{\Lambda ^{4} }{ 4-n \tilde{r}} \,   \qquad &  (n>2) 
	\end{cases}
\end{align}
Interestingly, for superrotations with $n=2$ the power spectrum can be rewritten in terms of the Stefan-Boltzman law but with an effective emitting area which diverges logarithmically
\begin{align}
 \qquad P_{SR}=\frac{1}{2}\sigma_s A_\text{eff} T^4\, , \quad  \quad A_\text{eff} = \frac{c^2+c^{-2}}{4 \pi} A \log \left( \frac{\Lambda}{T} \right )\,.
\end{align} 
For superrotations symmetric between the two poles the effective area further simplifies to $A_\text{eff} =  A \log(\Lambda/T)/4\pi$. The  $4 \pi$ difference to the Hawking case reflects the fact that in the case of superrotations the spectrum is dominated by the region near the poles and so there is an area suppression.
Regarding the logarithimic enhancement, let us provide an order-of-magnitude estimate to show that it might not be that dangerous after all. For example, if we fix the most extreme values of $\Lambda= M_p$ and a pick a black hole of the size of the present causal horizon, the enhancement in the spectrum would only be $\log (M_p/10^{-33}\text{eV}) \sim 10^2 $. Logarithmic corrections to the black hole area-law have also been obtained in other contexts \cite{Carlip:2000nv,Ghosh:2004rq}.

On the other hand, the total integrated power exhibits divergences in the UV between linear, for $n=3$, and quadratic, for $n \rightarrow +\infty$. 
A possible way to produce a physically more sensible result in these cases could be to regulate the states by following the methods of \cite{Compere:2018ylh,Donnay:2020guq} and add appropriate counter-terms at $\IP$ and $\IM$. This, however, goes beyond the scope of this paper. On the other hand, we note that modes with $m \neq 0$ do not give divergent contributions. The singular behaviors only happen for configurations with $m=0$, those with a nonzero profile at the poles, 
and are probably related to the presence of a cosmic string along the same axis, as proposed in \cite{Strominger2017}.

\section{Conclusions and Outlook}

In this work we discussed how the Hawking spectrum can be affected by supertranslations and superrotations. These transformations comprise asymptotic symmetries of asymptotically flat spacetimes and, as such, have been proposed as rightful symmetries of the gravitational $S$ matrix. 
In particular, such transformations do not spoil the orthogonality between different one-particle states and leave the space of asymptotic solutions of the wave equation invariant.
Our approach to explore their consequences on Hawking radiation has been to consider supertranslated or superrotated asymptotic states for a massless minimally coupled scalar field, while leaving the underlying dynamics of the spherically symmetric gravitational collapse, and in particular the relation between advanced and retarded time, unchanged.

We started by revisiting Hawking's derivation \cite{Hawking1975} and by looking at asymptotic states which are subject to supertranslations. 
We obtained in \eqref{mainST} and \eqref{bbST1} nontrivial off-diagonal phases in the two-point function as a consequence of the interplay between the transmission coefficient and the supertranslation function at $\mathscr I^+$. These phases become negligible at very high energies, where the transmission coefficient becomes unity, but they cannot be neglected in intermediate regimes where the frequency is comparable to the height of the potential barrier generated by the angular momentum.
The emitted particle number and power spectrum, associated with the diagonal entries, eventually remain unmodified as already anticipated by Hawking \cite{Hawking1975} and further elaborated upon in \cite{Javadinezhad2019,Compere2019}. 

We then moved to the case of boosted and superrotated states where we found more substantial differences. We obtained that, compared to the original case, the frequency and direction dependence of the two point function are corrected, respectively, by the conformal factor  $\omega \mapsto \omega R_+(\theta,\phi)$ and by the superrotation transformation $\hat q \mapsto G_+(\hat q)$ at  $\IP$ (see in particular \eqref{bbSR qbasis}). Instead, the corresponding transformation at $\IM$ does not affect the spectrum. This  correction, which encodes the direction-dependent  red/blueshift of the frequency due to the boost or the superrotation, is the source of all the subsequent results. First, it generates correlations between modes with different angular quantum numbers, which is natural given the angle dependence of the transformation. Second, it also corrects the frequency dependence of the spectrum. 
While in the case of boosts the correction to the particle emission spectrum does not affect the exponential suppression for large enough frequencies, that is no longer the case for superrotations. Interestingly, for the latter, the corresponding integral over the celestial sphere is finite and receives its leading contributions from angles $\theta$ such that $\omega R_+\simeq1$, even though these transformations are singular at the two poles of the sphere. As a result, the spectral rate of particle emission exhibits a characteristic power-law decay at high frequency.
Analogous features are also shared by ultrarelativistic boosts.

Afterwards, we explored the impact on the total power and total particle emission rate. For a boosted state these quantities are rescaled with respect to the standard, static ones in accordance with expectations due to the Doppler effect. However, for the superrotated states the power-law behavior inherited by the particle number generically leads to power-law divergences in these integrated quantities. The origin of these divergences are the modes of the field with quantum number $m=0$, those with a non-zero profile at the singular points of the superrotation transformation. These singular points have been associated with the presence of a cosmic string and so such divergences might be a further signature of that.
We found, however, an interesting exception, the particular case where the superrotation corresponds to a double-cover transformation $(z,\bar z)\mapsto c (z^2, \bar z^2)$. In this case the total particle rate is finite while the total power can be shown to still satisfy the Stefan-Boltzmann law but with an rescaled effective area which diverges logarithmically in the ultraviolet (see \eqref{power spectrum superrotations}).

The results of this work open several interesting avenues for future work to which we plan to return. For example, in \eqref{mainST} and \eqref{mainSR} we found that the spectrum is not diagonal in the angular quantum numbers. It would be interesting to explore further consequences of these correlations induced by supertranslations and superrotations, in particular, understand if the off-diagonal entries which we found for supertranslations can leave any imprint in measurable quantities. Moreover, the full quantum state at $\IP$ is not solely described by its two-point function so it would be important to study how the supertranslated and superrotated states affect higher-point correlators of the form $\langle (\boldsymbol{b}^\dagger)^m (\boldsymbol{b})^n\rangle$ and whether they can provide additional information about the black hole. Generalizing the calculation to particles with spin is another interesting route as we expect more non-trivial angular features in those cases.
	
Another relevant question concerns the generalization of the present derivation to the case of the extended superrotations first proposed by Campiglia and Laddha \cite{Campiglia:2014yka}, where instead of conformal transformations one deals with the group of diffeomorphisms on the sphere.
These transformations are supposed to be smoother, compared to conformal transformations, and could ameliorate some of the shortcomings of the present discussion. For instance, one could envision a diffeomorphism which is akin to a conformal transformation with $R \ll 1$ near the poles, while still remaining nonsingular. 	
On ther other hand, it would also be interesting to understand if the divergences found for most superrotations in the total power emitted could be regulated by appropriately adding boundary counter-terms, for example, by following the procedure in \cite{Compere:2018ylh,Donnay:2020guq}.

From an asymptotic-symmetry perspective, a natural problem would be to identify possible relations between the asymptotic symmetry algebra and the modified spectra, exploring in particular how the commutator of two asymptotic symmetries is reflected in the resulting radiation. 

Another interesting point concerns the derivation of Hawking radiation proposed by Damour and Ruffini \cite{Damour:1976jd}, where the black-hole temperature is derived from an analytic continuation of an outgoing wave solution inside the black-hole horizon, and the asymptotic properties of the states at $\mathscr I^\pm$ seemingly play little role. 
A natural attempt to re-discuss the effect of asymptotic symmetries on the emission spectra would be to study appropriate supertranslations/superrotations defined at the horizon itself (see e.g. \cite{Donnay:2015abr,Donnay:2016ejv}). 
	
	 The most physically compelling and conceptually challenging question is however to which extent the effects of superrotations on the spectrum can shed light, if at all, on the problem of information loss. 
	The first steps in this direction would entail having a clear understanding of the physicality of the off-diagonal entries found for supertranslations and of the superrotated charges, for example, by exploring whether they could be associated with a dynamical process without relying on the insertion of a cosmic string. The other natural step would be to calculate the corresponding contribution to the black-hole entropy and identify a suitable microstate counting procedure to retrieve it from an underlying microscopic description.

\subsection*{Acknowledgements}
We would like to thank Paolo Di Vecchia and Dario Francia for useful comments on a preliminary version of this work. The work of C.H. is supported by the Knut and Alice Wallenberg Foundation under grant KAW 2018.011.

\appendix

\section{Quantization and Wave Equation}
\label{app:basics}

\subsection{Quantization of the scalar field}\label{app:quantization-scalar}
Let us summarize here the main steps involved in the quantization of a massless Hermitian scalar field on a general curved background (see e.g.~\cite{Birrell:1982ix}). Given a local observer, one considers an orthonormal basis of \emph{positive-frequency} solutions $f_i$ of the scalar wave equation
\begin{equation}\label{BoxM}
\Box f = g^{ab}\nabla_a \nabla_b f=0\,. 
\end{equation} 
Here the notion of positive frequency is defined with respect to the observer's local time, while orthonormality is defined with respect to the invariant scalar product
\begin{equation}
\left(f, g \right) = \frac{i}{2}\int_\Sigma \left(f\, \partial_a \bar g - \bar g\, \partial_a f\right) \, d\Sigma^a\,,
\end{equation}
$\Sigma$ being any Cauchy surface, so that
\begin{equation}\label{orthonormal}
(f_i, f_j)=\delta_{ij}\,,\qquad
(f_i,\bar f_j)=0\,,\qquad (\bar f_i, \bar f_j)=-\delta_{ij}\,.
\end{equation} 
Note that $(f,g)=\overline{(g,f)}=-(\bar g, \bar f)$.  
The functions $f_i$ represent the states of free particles as seen by the local observer.
 The quantized Klein--Gordon field is then defined by the expansion
\beq
\boldsymbol{\phi}=\sum_{i}\left\{f_{i}\, \boldsymbol{a}_{i}+\bar{f}_{i}\, \boldsymbol{a}_{i}^{\dagger}\right\}
\eeq
where the operators $\boldsymbol{a}_{i}$ and $\boldsymbol{a}_{i}^\dagger$ satisfy
\beq \label{commutator}
[\boldsymbol{a}_{i},\boldsymbol{a}_{j}^\dagger] = \delta_{ij}\,,\qquad [\boldsymbol{a}_{i},\boldsymbol{a}_{j}] = 0\,.
\eeq
The Fock vacuum $|0_-\rangle$ is defined by requiring that
$\boldsymbol{a}_{i}|0_-\rangle = 0$ for all $i$.

A different observer will in general have inequivalent notions of local time and positive frequency. Accordingly, their preferred orthonormal basis of single-particle states $p_i$ will in general differ from the one discussed above. The corresponding expansion for the Klein--Gordon field then reads
\beq
\boldsymbol{\phi}=\sum_{i}\left\{p_{i}\, \boldsymbol{b}_{i}+\bar{p}_{i}\, \boldsymbol{b}_{i}^{\dagger}\right\}\,,
\eeq
where $\boldsymbol{b}_{i}$ and $\boldsymbol{b}_{i}^\dagger$ satisfy commutation relations analogous to \eqref{commutator}, and the Fock vacuum is identified by demanding that
$\boldsymbol{b}_{i}|0_+\rangle = 0$ for all $i$.

The relation between the two choices $f_i$ and $p_i$ is characterized by the Bogolyubov coefficients
\begin{equation}\label{}
\alpha_{ij} = (p_i,f_j)\,,\qquad
\beta_{ij} = - (p_i,\bar f_j)\,,
\end{equation}
which grant
\begin{equation}\label{}
p_i = \sum_j \left(\alpha_{ij}f_j + \beta_{ij}\bar f_j\right) \,,\qquad
f_i = \sum_j \left(\bar \alpha_{ji}\, p_j - \beta_{ji}\,\bar p_j\right)\,,
\end{equation}
and hence
\begin{equation}\label{key}
\boldsymbol{b}_{i} = \sum_j \left(\bar \alpha_{ij}\boldsymbol{a}_{j} - \bar \beta_{ij} \boldsymbol{a}_{j}^{\dagger}\right)\,.
\end{equation}
The number of particles measured by the second observer in the vacuum defined by the first observer is therefore encoded in the diagonal entries of the correlator
\begin{equation}\label{}
\langle 0_-| \boldsymbol{b}_{i}^\dagger \boldsymbol{b}_{j} |0_-\rangle = \sum_k \beta_{ik} \, \bar \beta_{jk}\,.  
\end{equation}
By consistency of the Bogolyubov coefficients with orthonormality, one also has
\begin{equation}\label{key}
\sum_{k}\left(\alpha_{ik} \, \bar \alpha_{jk}-\beta_{ik} \, \bar \beta_{jk}\right)= \delta_{ij}\,.
\end{equation}

In the body of the text we will often be dealing with the surfaces $\mathscr I^\pm$. Considering two solutions whose asymptotics at $\mathscr I^-$ are 
\begin{equation}\label{}
	f(v,r,\hat x) \,\sim\, \frac{1}{r}\, f^{(1)}(v,\hat x)\,,\qquad g(v,r,\hat x) \,\sim\, \frac{1}{r}\, g^{(1)}(v,\hat x)\,,
\end{equation}
then their scalar product can be evaluated on $\mathscr I^-$ obtaining (in four spacetime dimensions)
\begin{equation}\label{SP}
	(f,g) = \frac{i}{2} \oint d\Omega(\hat x) \int_{-\infty}^{+\infty}dv \left[f^{(1)}(v,\hat x)\partial_v \bar g^{(1)}(v,\hat x)-\bar g^{(1)}(v,\hat x)\partial_vf^{(1)}(v,\hat x)\right]
\end{equation}
where $d\Omega$ is the usual measure on the Euclidean two-sphere.
Whenever the integral over $v$ converges, the scalar product can also be cast in the form
\begin{equation}\label{IBP}
	(f,g) = {i} \oint d\Omega(\hat x) \int_{-\infty}^{+\infty}dv \, f^{(1)}(v,\hat x)\partial_v \bar g^{(1)}(v,\hat x)
\end{equation}
using integration by parts. Similar relations hold at $\mathscr I^+$, given the appropriate asymptotics.

\subsection{The wave equation in flat spacetime}\label{app:wave-flat}
In flat spacetime, there exist global notions of time and positive frequency. A commonly adopted choice as the basis $f_i$ is given by plane waves,
\beq\label{planewave}
f_i \longmapsto f_{\mathbf k}(x) = \tfrac{1}{(2\pi)^{\frac{D-1}{2}}\sqrt{|\mathbf k|}}\,  e^{i |\mathbf k| t-i\mathbf k\cdot\mathbf  x}\,.
\eeq
Let us also recall the solution of the flat--spacetime wave equation in spherical coordinates. We thus introduce coordinates such that
$
\mathbf x = r\,\hat x\,,
$
where $\hat x$ is a unit vector parametrized by $D-2$ angles $\xi^A$. In these coordinates, the wave equation takes the form
\beq
\Box f = -\partial_t^2 f + \frac{1}{r^{D-2}}\partial_r\left(r^{D-2}\partial_r f\right)+\frac{1}{r^2}\Delta_{D-2}f\,,
\eeq
where $\Delta_{D-2}$ is the Laplacian on the unit sphere. The ansatz
\beq
f_{\omega \ell} = \frac{e^{i\omega t}}{\sqrt{2\pi \omega}}\, \frac{1}{r^{\frac{D-2}{2}}}\,F_{\omega l}(r)\, Y_\ell(\hat x)\,,
\eeq
with $Y_\ell(\hat x)$ the hyperspherical harmonics ($\ell$ is a suitable collection of indices, e.g. $\ell=(l,m)$ in $D=4$), reduces the wave equation to
\beq\label{Besseleq}
F''_{\omega l}(r)+\left[
\omega^2-\frac{\left(\tfrac{D}{2}-1\right)\left(\tfrac{D}{2}-2\right)}{r^2}-\frac{l(l+D-3)}{r^2}
\right]F_{\omega l}(r)=0\,.
\eeq 
For large radius $r$, this equation reduces approximately to 
\beq
F''_{\omega l}(r)+
\omega^2F_{\omega l}(r)\approx0
\eeq
so that it admits the two independent asymptotic behaviors
$e^{\pm i\omega r}$,
which correspond to incoming or outgoing waves 
\beq
f_{\omega \ell} \sim \frac{e^{i\omega (t\pm r)}}{\sqrt{2\pi\omega}}\, \frac{1}{r^{\frac{D-2}{2}}}\, Y_\ell(\hat x)\quad\qquad (r\to\infty)\,.
\eeq
The general solution of \eqref{Besseleq} can be cast in terms of Bessel functions of the third kind by 
\beq
F_{\omega l}(r) = c_1 \sqrt{\omega r}\, H^{(1)}_{\nu}(\omega r) + c_2 \sqrt{\omega r}\, H^{(2)}_\nu (\omega r)\,,
\eeq
where
$
2\nu^2-1 =  \left(D-2\right)\left({D}-4\right) + 4l(l+D-3)
$.
Indeed, these functions have the asymptotic form
\beq
\sqrt{\omega r}\, H^{(1)}_{\nu}(\omega r) \sim \sqrt{\frac{2}{\pi}} \,e^{i\left(\omega r - \frac{1}{2}\nu\pi-\frac{\pi}{4}\right)}\,,\qquad
\sqrt{\omega r}\, H^{(2)}_{\nu}(\omega r) \sim \sqrt{\frac{2}{\pi}} \,e^{-i\left(\omega r - \frac{1}{2}\nu\pi-\frac{\pi}{4}\right)}\,,
\eeq
as $r\to\infty$. Thus, 
\beq
f_{\omega \ell} = c_1\, \frac{e^{i\omega t}}{\sqrt{2\pi \omega}}\,\frac{\sqrt{\omega r}}{r^{\frac{D-2}{2}}}\, H_\nu^{(1)}(\omega r) Y_\ell(\hat x)+c_2\, \frac{e^{i\omega t}}{\sqrt{2\pi \omega}}\,\frac{\sqrt{\omega r}}{r^{\frac{D-2}{2}}}\, H_\nu^{(2)}(\omega r) Y_\ell(\hat x)
\eeq
is the general solution to the wave equation.

\subsection{The wave equation in Schwarzschild spacetime}\label{app:wave-Schwarzschild}

The metric of Schwarzschild spacetime can be written as follows
\beq\label{Schwtr}
ds^2 = - F(r)\, dt^2 + \frac{dr^2}{F(r)} + r^2 d\Omega^2_{D-2}\,,
\qquad F(r) = 1-\left(\frac{r_s}{r}\right)^{D-3}\,.
\eeq
The coordinate $t$ is the Schwarzschild time, the proper time of a static observer far away from $r_s$, while $r$ has the meaning of a luminosity distance.
It is convenient to introduce the ``tortoise'' coordinate $r_\ast$ by means of the differential equation
\beq\label{tortoise}
\frac{dr_\ast}{dr} = \frac{1}{F}\,.
\eeq
The explicit solution of \eqref{tortoise} is elementary in $D=4$ and yields
\beq\label{tortD=4}
r_\ast = r + r_s \log \left( \frac{r}{r_s}-1\right) + {C}
\eeq
for $r>r_s$
where $C$ is customarily set to zero.
The metric then reads
\beq\label{Schwtrast}
ds^2 = F \left( -dt^2 + dr_\ast^2 \right) + r^2 d\Omega_{D-2}^2
\eeq
and the wave equation can be written in the form
\beq
-\partial_t^2 f + \frac{1}{r^{D-2}}\partial_{r_\ast}\left(r^{D-2} \partial_{r_\ast} f \right) + \frac{F}{r^2}\,\Delta_{D-2}f=0\,.
\eeq
It should be recalled that $r=r(r_\ast)$ and $F=F(r(r_\ast))$.
Steps analogous to those employed in Minkowski spacetime lead to the basis of solutions
\beq \label{solution for spherically symmetric space-times}
f_{\omega \ell} = e^{i\omega t}\,\frac{1}{r^{\frac{D-2}{2}}}\, F_{\omega l}(r_\ast)\,Y_\ell(\hat x)\,,
\eeq
where the radial function $F_{\omega l}(r_\ast)$ must now satisfy a deformation of the Bessel equation:
\begin{align}
	F''_{\omega l}(r_\ast)+ \omega_\text{eff}^2 \, F_{\omega l}(r_\ast) =0  \, ,
\end{align}
where 
\begin{align}
\omega_\text{eff}^2=  \omega^2-F\left[\frac{\left(\tfrac{D}{2}-1\right)\left(\tfrac{D}{2}-2\right)F}{r^2}+\dfrac{(D-2)(D-3)}{2}\frac{r_s^{D-3}}{r^{D-1}}+\frac{l(l+D-3)}{r^2}\right] \, .
\end{align}
Although the solutions of this equation do not admit a simple closed form, inspecting at the large--$r$ limit one sees that their asymptotic behavior must be $e^{\pm i\omega r_\ast}$, thus identifying ingoing and outgoing waves \cite{Birrell:1982ix}. 

\section{Spherical Harmonics \label{app: spherical harmonics}}
Here we list a few useful expansions for the spherical harmonics and some related functions.
We find it convenient to label points on the two sphere either by a pair of angles, e.g. $\xi=(\theta,\phi)$ in standard polar and azimuthal coordinates, or by unit vectors $\hat x$. 
The spherical harmonics $Y_{lm}(\theta,\phi)$ are defined by
\begin{align} \label{spherical harmonic expansion}
	Y_{lm}(\theta, \phi) = s_{lm}  P_{l}^m(\cos \theta) e^{im \phi} \, , \qquad s_{lm} \equiv (-1)^m \sqrt{\frac{2l+1}{4\pi} \frac{(l-m)!}{(l+m)!}}
\end{align}
where $P_l^m(x)$ are the associated Legendre polynomials. 
These functions are eigenstates of the orbital angular momentum and satisfy the orthogonality and completeness conditions
\begin{equation}\label{}
\oint d\Omega(\hat x) Y_{lm}(\hat x)Y_{l'm'}(\hat x) = \delta_{ll'} \delta_{mm'}\,,\qquad
\sum_{lm} Y_{lm}(\hat x) Y_{lm}(\hat x') = \delta(\hat x, \hat x')
\end{equation}
where $\delta(\hat x, \hat x')$ is the invariant delta function on the two-sphere.

We will be interested in expanding the spherical harmonics around the south and north pole of the sphere ($\theta=0,\pi$, corresponding to $x=\cos\theta=\pm 1$). 
Then, for integer $m$ we can use
\begin{equation} \label{Legendre polynomial near the poles}
	P_l^m(x) \,\sim\,  p_{lm} \left(\frac{1-x}{2}\right)^{|m|/2}\qquad (x\rightarrow 1^-)
\end{equation}
where, letting  $(a)_k$ denote the Pochhammer symbol,
\begin{equation}
	p_{lm}=
	\begin{cases}
		\dfrac{1}{\Gamma(1-m)} \qquad & m=0,-1,-2, \ldots \\ \\
		\dfrac{(-1)^m }{m!}  (l-m+1)_{2m} \qquad &m=1,2, \ldots
	\end{cases}
\end{equation}
and similarly for $x\rightarrow -1^+$ following from the relation
\begin{equation} \label{reflection law}
	P_l^m(-x)= (-1)^{l-m} P_l^m(x) \, .
\end{equation}
Therefore the following limit defines a finite quantity
\begin{equation} \label{cpm}
	c_{ll'm}^\pm\equiv\lim_{x\to\pm 1}\frac{P_l^m(x)P_{l'}^m(x)}{(1-|x|)^{|m|}} \,,
\end{equation}
and we find, for any integer $m$,
\begin{equation}
	c_{ll'm}^- = (-1)^{l+l'} c_{ll'm}^+\, .
\end{equation}

Let us also recall here the addition theorem
\begin{equation}\label{YYPx}
	\sum_{m}Y_{lm}(\hat x)Y_{lm}^\ast(\hat x') = \frac{2l+1}{4\pi} \, P_l(\hat x \cdot \hat x')
\end{equation}
where $P_l(x)$ is the $l$th Legendre polynomial. Since $P_l(1)=1$, a special case of this result is the identity
\begin{equation}\label{YYP}
	\sum_{m}|Y_{lm}(\hat x)|^2 = \frac{2l+1}{4\pi}\,.
\end{equation}

\section{Geometry of Spherical Collapse \label{app: Geometry of spherical collapse}}
We consider a spherically symmetric, homogeneous matter distribution---a ``star''---that undergoes a collapse. Following in particular \cite{Adler:2005vn}, 
we assume that metric in the interior of the star is given by a Friedmann--Robertson--Walker metric
\beq\label{FRWtcrc}
ds^2 = -dt_c^2 + a(t_c)^2 \left[
dr_c^2 + r_c^2 d\Omega_{D-2}^2
\right]\,,
\eeq 
where $t_c$ is the co--moving time, while the co--moving radius $r_c$ is truncated at a value $\bar r_c$ identifying the boundary of the star itself,
$0\le r_c \le \bar r_c$.
The conformal time $\eta$, defined by 
\beq
{dt_c} = {a(t_c)}\, {d\eta}\,,\qquad \eta\big|_{t_c=0} = 0\,,
\eeq
brings the metric to the form
\beq\label{FRWetarc}
ds^2 = a^2 \left[-d\eta^2 + 
dr_c^2 + r_c^2 d\Omega_{D-2}^2
\right]
\eeq
and is such that, for null rays traveling at fixed angles,
$
d\eta= \pm dr_c\,.
$
Introducing the physical radius (luminosity distance) by 
$
r = a(t_c) r_c
$
we find
\beq\label{FRWtcr}
ds^2 = -\left[1-\left(\frac{r\dot a}{a}\right)^2\right]dt_c^2 - 2\left(\frac{r\dot a}{a}\right)dt_c\, dr + dr^2 + r^2 d\Omega_{D-2}^2\,.
\eeq
On the exterior of the star, the metric takes instead the Schwarzschild form \eqref{Schwtr}.
In order to identify the proper matching with \eqref{FRWtcr}, we introduce a new time coordinate $t_c$ (in principle different from the co-moving time $t_c$ only defined inside the collapsing matter) according to 
\beq\label{ttc}
t = t_c + g(r)\,.
\eeq  
Substituting, we see that the metric is brought to the form
\beq\label{Schwtcr}
ds^2 = -F(r)dt_c^2 + 2\sqrt{1-F(r)}\,dt_c dr + dr^2 + r^2 d\Omega_{D-2}^2\,,
\eeq
provided 
\beq
g'(r) = - \, \frac{\sqrt{1-F(r)}}{F(r)}\,.
\eeq
A solution of this equation in $D=4$ is given by 
\beq\label{ttcD=4}
g(r) = - 2\sqrt{r_sr}-r_s \log \left(\frac{\sqrt{r}-\sqrt{r_s}}{\sqrt{r}+\sqrt{r_s}}\right)\,.
\eeq
Comparing \eqref{FRWtcr} and \eqref{Schwtcr}, we see that $t_c$ is indeed the co--moving time and that the correct matching between the two expressions for the metric is 
\beq
\frac{\dot a(t_c)}{a(t_c)}= - \frac{\sqrt{1-F(r)}}{r}\Big|_{r = R(t_c)}\,,
\eeq 
where $R(t_c)=a(t_c) \bar r_c$ is the physical radius at the boundary of the star.

Consider now a future--directed light ray leaving from the origin $r_c=0$ at a co--moving time $t_c=t_0$ in a given direction. Its trajectory satisfies
$
d\eta = dr_c
$
namely
\beq
\frac{dr}{dt_c} = 1 + \frac{r\dot a}{a}\,,\qquad r\big|_{t_c=t_0} =0
\eeq
or, equivalently,
\beq
r(t_c) = t_c-t_0 + \int_{a(t_0)}^{a(t_c)} r\,\frac{da}{a}\,.
\eeq
Assuming that the emission has occurred before the formation of the event horizon, this ray will eventually emerge from the collapsing body at a co--moving time $T_+$ dictated by 
\beq
R(T_+) = T_+ - t_0 + \int_{a(t_0)}^{a(T_+)} r\,\frac{da}{a}\,.
\eeq
The retarded time coordinate for the Schwarzschild geometry is defined by $u = t-r_\ast$ and thus, recalling \eqref{ttc},
the null ray emerges from the star at a retarded time 
\beq
u = T_+ + g(R(T_+))-r_\ast(R(T_+))\,.
\eeq
Considering instead a null ray arriving in the origin at the co-moving time $t_0$, we can retrace the above steps, choosing in particular $d\eta= -dr_c$, to conclude that it had entered the collapsing star at the co--moving time $T_-$ defined by 
\beq
R(T_-) = t_0 - T_- + \int_{a(t_0)}^{a(T_-)} r\,\frac{da}{a}\,,
\eeq
and at the advanced time
\beq
v = T_- + g(R(T_-))+r_\ast(R(T_-))\,.
\eeq
Combining \eqref{tortoise} and \eqref{ttc} we may write
\beq
u = T_+ + g_-(R(T_+))\,,\qquad
v = T_- + g_+(R(T_-))\,,
\eeq
where
\beq
g_\pm(r) = g(r)\pm r_\ast(r)
\eeq
satisfy the differential equation
\beq\label{gpmdiff}
\frac{dg_\pm}{dr} = \frac{-\left(\tfrac{r_s}{r}\right)^{\frac{D-3}{2}}\pm 1}{1-\left(\tfrac{r_s}{r}\right)^{D-3}}\,.
\eeq
The near--horizon behavior of $g_-$, namely for $0<1-\frac{r_s}{r}\ll 1$, is thus captured by
\beq
\frac{dg_-}{dr} \approx -\frac{2}{(D-3)\left(1-\tfrac{r_s}{r}\right)}\,,
\eeq
which integrates to
\beq
g_- \approx -\frac{2}{D-3}\left(r + r_s \log\left(\frac{r}{r_s}-1\right)\right)+C\,.
\eeq
On the other hand, we see that $g_+$ is regular near the horizon and satisfies $g_+\approx \frac{r}{2}+C$.

We now assume for simplicity that the scale factor $a$ does not vary appreciably during the time our light ray spends inside the star (this assumption can be generalized without spoiling the final outcome \cite{Ford:2001gp}), obtaining the approximate expressions
\beq
R(T_+) \approx T_+-t_0\,,\qquad
R(T_-) \approx t_0-T_-\,
\eeq
from which we can eliminate $t_0$ to obtain a quantity that does not depend on the specific time at which the ray crosses the origin
\beq
R(T_+) + R(T_-)\approx T_+-T_-\,.
\eeq
Consider now a co--moving time $t_0$ in the limit in which the light ray emerges at the horizon, so
\beq
R(T_+)\to r_s\,.
\eeq
We see that, by the asymptotics of $g_-$, this corresponds to an infinite retarded time
\beq
u\to+\infty
\eeq
while the light had entered the star at the (finite) advanced time
\beq
v \to T_-+g_+(T_+-T_--r_s) \equiv v_0\,.
\eeq
For a light ray emerging slightly before the horizon formation, instead,
\beq
R(T_+)= r_s(1+\epsilon)\,,
\eeq
for small positive $\epsilon$, we have
\beq
u\approx -\frac{2r_s}{D-3} \log \left({\epsilon}\right)\,, 
\eeq
up to finite terms in the limit $\epsilon\to0$,
while correspondingly
\beq\begin{aligned}
	v &= T_-+g_+(T_+-T_--r_s(1+\epsilon)) \\
	&\approx v_0 - g_+'(T_+-T_--r_s)  r_s \epsilon\,,
\end{aligned}\eeq
or using \eqref{gpmdiff}
\beq
v_0-v \approx  \frac{1-\left(\tfrac{r_s}{T_+-T_--r_s}\right)^{\frac{D-3}{2}}}{1-\left(\tfrac{r_s}{T_+-T_--r_s}\right)^{D-3}}\, r_s\epsilon.
\eeq
We may absorb the positive coefficient multiplying $\epsilon$ in a phenomenological constant $C$ that depends on the details of the collapse, encoded in our case in the specific values of $T_+$, $T_-$.
In conclusion, under the simplifying assumption that the variation of the scale factor describing the collapsing star is negligible, a light wave exiting the star at a retarded time $u$, slightly before the horizon formation, had entered it at an advanced time $v<v_0$ given by
\beq
u \approx -\frac{2r_s}{D-3} \log \frac{v_0-v}{C}\,,
\eeq
where $C$ is a phenomenological parameter. In particular, the coefficient appearing on the right--hand side is related to the surface gravity $\kappa$ of the black hole, given by $2\kappa = F'(r_s)$, 
so that
\beq
\frac{2r_s}{D-3}=\frac{1}{\kappa}\,. 
\eeq
All the above formulas can be verified more directly in dimension four, using the simple solutions \eqref{tortD=4} and \eqref{ttcD=4}. In particular
$
2r_s = 4M = \kappa^{-1}
$
in $D=4$,

\section{Conformal Transformations}
\label{app:conformal}

The mapping $\xi'^A=G^A(\xi)$ is a conformal transformation if the transformed metric is equal to the original metric $\gamma_{AB}$ up to a an overall positive factor $R(\xi)$ called conformal factor,
\begin{equation}\label{defconfmapp}
\gamma'_{AB}(\xi')=
\frac{\gamma_{AB}(\xi')}{R^2(\xi)}\,.
\end{equation}
In two dimensions, this relation implies that, for a generic function $f$, 
\begin{equation}\label{confintegral}
\int d^2\xi \sqrt{\gamma(\xi)} R^2(\xi) f(G(\xi)) = \int d^2\xi' \sqrt{\gamma(\xi')} f(\xi')\,, 
\end{equation}
where $\gamma$ stands for the determinant of $\gamma_{AB}$.

Local conformal transformations on the Euclidean two-sphere take a particularly simple form in stereographic coordinates
\begin{equation}\label{}
z=e^{i\phi}\tan\frac{\theta}{2}\,,
\end{equation}  
in terms of which the metric on the sphere reads
\begin{equation}\label{dzdzbar}
ds^2 = \frac{4 dz d\bar z}{(1+z\bar z)^2}\,.
\end{equation}
They are given by holomorphic and anti-holomorphic transformations,
\begin{equation}\label{}
(z,\bar z)\mapsto (z', \bar z\,')=(G(z), \tilde G(\bar z))\,.
\end{equation}
Performing the change of variables indeed yields
\begin{equation}\label{}
ds^2 = \frac{4 dz' d\bar z\,'}{(1+z'\bar z\,')^2}\,\frac{(1+G(z)\tilde G(\bar z))^2}{G'(z)\tilde G'(\bar z)(1+z\bar z)^2}\,,
\end{equation}
where we have isolated the old metric written in the new coordinates in the first factor.
By definition of conformal factor \eqref{defconfmapp}, we then have
\begin{equation}\label{Rsquate for a general transformation}
R^2(z,\bar z) = \frac{G'(z)\tilde G'(\bar z)(1+z\bar z)^2}{(1+G(z)\tilde G(\bar z))^2}\,.
\end{equation}

As described in Section \ref{sec:BMSreview}, local conformal transformations naturally emerge in the discussion of the BMS group.
Except for global $SL(2,\mathbb C)$ transformations 
which correspond to the action of the Lorentz group (boost and rotations), other general holomorphic transformations will have branch cuts and/or divergences \cite{Adjei:2019tuj,Oblak:2015qia}.

We now restrict out attention to two special classes of conformal transformations that arise from boosts and superrotations  via \eqref{SR} i.e. transformations of the type
\begin{equation}\label{}
G(z) = c\,z^n\,,\qquad
\tilde G(\bar z) =c\, \bar z^{n}\,,
\qquad n=1,2,3,4,\ldots
\end{equation}
which yield the conformal factor
\begin{equation}
R^2(z,\bar z) =(c\, n)^2 \frac{|z|^{2(n-1)}(1+|z|^2)^2}{(1+c^2|z|^{2n})^2}\,.
\end{equation}
In terms of the coordinates defined in \eqref{SR+1}, \eqref{SR+2} we thus obtain
\begin{equation}
R(z,\bar z)
 =  
\frac{n}{2}\, c^{\frac{1}{n}}(1+x)^{\frac{n+1}{2n
}}\left(1-x\right)^{\frac{n-1}{2n}}
\left[
1+\frac{1}{c^{\frac{2}{n}}}\left(\frac{1-x}{1+x}\right)^{\frac{1}{n}}
\right]\,,
\end{equation} 
where $\theta_+$ is the polar coordinate in the new coordinate system and $x=\cos(\theta_+)$. The case $n=1$ corresponds to boosts in the third spatial direction and the conformal factor simplifies to
\begin{equation}
\text{\textit{Boosts}:} \qquad   \qquad R_+ 
= \frac c 2  (x+1)+ \frac 1 {2c} (1-x)  \, .
\end{equation}
Note that for boosts $R_+ >0$ as long as the rapidity $c$ is finite and nonvanishing, while it approaches zero either at the north or at the south pole for ultrarelativistic boosts.
For $n=2,3,\ldots$ we have  
superrotations whose conformal factors asymptote to zero near the poles of the sphere as 
\begin{equation}
\begin{aligned}
\text{\textit{Superrotations}:} \qquad    R_+ \,&\sim\,  {n} \begin{cases}
c^{\frac{1}{n}}\left(\frac{1-x}{2}\right)^{\frac{n-1}{2n}}\qquad  (x\to 1^-)\\
c^{-\frac{1}{n}}\left(\frac{1+x}{2}\right)^{\frac{n-1}{2n}}\qquad  (x\to -1^+) \, .
\end{cases} 
\end{aligned}
\end{equation}
A $c\neq1$ gives rise to an asymmetry between northern  and southern hemisphere. 

\section{Integral in the Vicinity of $R=0$ \label{app:R=0}}

In the main text we found at different instances  a $\delta$-function of the form
\begin{equation}
\delta\left((\omega-\omega')R(\xi)\right) \, .
\end{equation}
and we neglected the zeros of $R(\xi)$. In this appendix we justify this approach. 

Let us start by looking at the regions where $\omega \neq \omega'$. In that case we would have
\begin{equation}
\delta((\omega-\omega')R(\xi)) = \frac{1}{\omega-\omega'} \sum_{\xi_\ast}  \frac{\delta(\xi,\xi_\ast)}{|\nabla R(\xi_\ast)|} \,,
\end{equation}
where $\xi_\ast$ are the zeros of $R(\xi)$.
For the class of superrotations considered in the body of the paper, we find $1/|\nabla R(\xi_\ast)| = 0$. In addition, the integrand multiplying the $\delta$-function vanishes at $\xi_\ast$  in all the relevant expressions. Therefore, these contributions vanish identically.

Therefore, the only place where the zeros of $R(\xi)$ could potentially contribute is when $(\omega - \omega')  \rightarrow 0$. To deal with this case we ignore temporarily the zeros of $R(\xi)$ and write
\begin{equation}
\delta((\omega-\omega')R(\xi)) \, =  \frac{1}{R(\xi)}\,\delta(\omega-\omega') \, .
\end{equation}
Then, we check if the results after integrating over the sphere receive any contribution from the zeros of $R(\xi)$. Again due to the observation that the integrand multiplying the $\delta$-function vanishes at the zeros $\xi_\ast$, we have explicitly obtained that in all instances the integrals do not receive contributions from such points, but rather from small regions around them.

\section{Transmission Coefficients and Density of States \label{app: Transmission coefficients}}

In this appendix we elaborate on the form of the transmission coefficients and the associated density of states $g(\omega)$ defined in \eqref{eq: density of states}. 
We will focus on $|t_{\omega l}|^2$, rather than $t_{\omega l}$, as all the results derived for the two point function and for the spectrum only depend in this quantity. We refer the reader to  \cite{Starobinsky:1973aij,Starobinskil:1974nkd,Page:1976df,Sanchez:1976fcl,Sanchez:1976xm,Sanchez:1977vz, Gray:2015xig} for a more thorough discussion.

\begin{figure}
	\centering
	\includegraphics[width=0.4\linewidth]{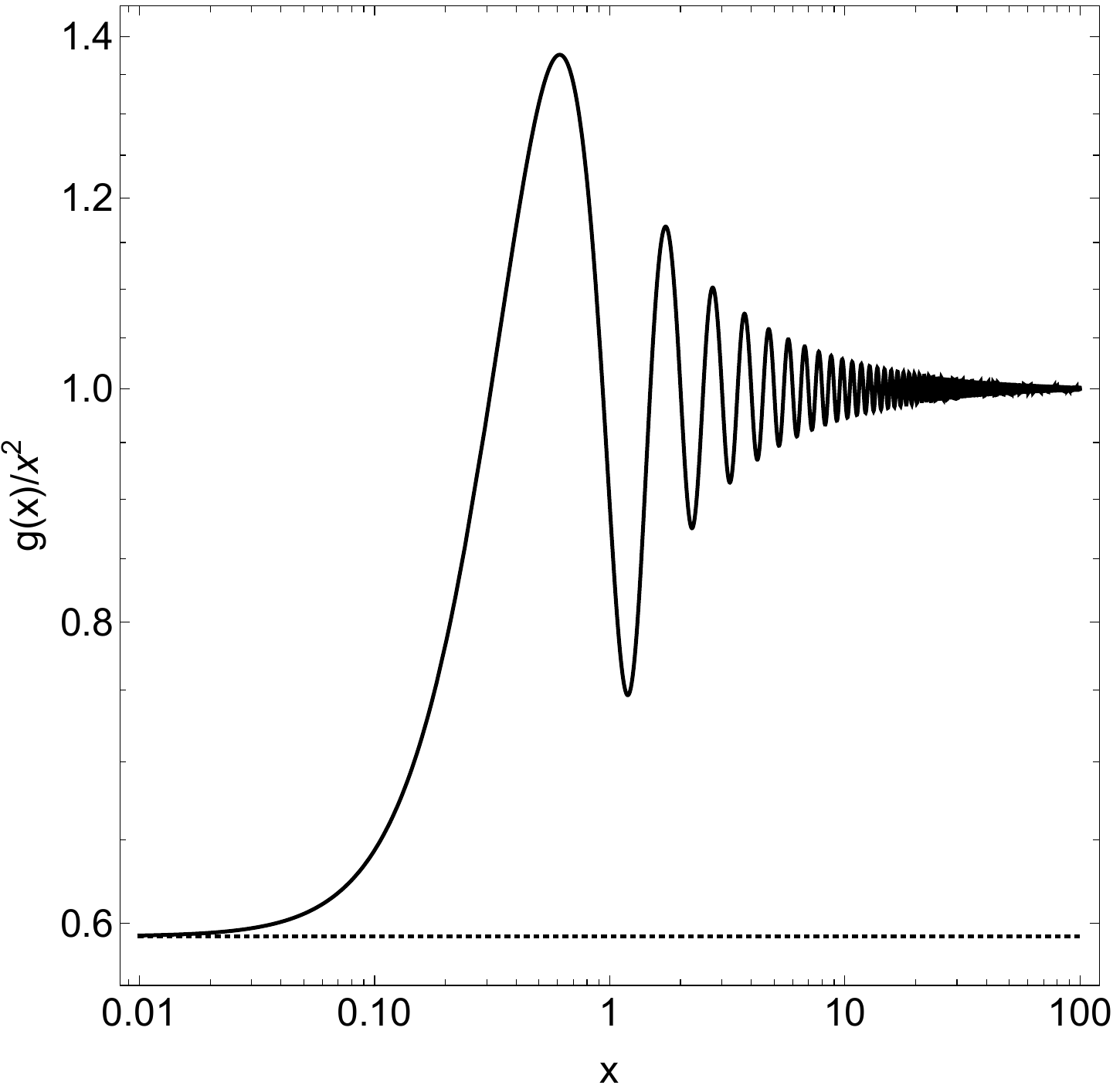}
	\caption{The rescaled density of states $g(x)/x^2$ with $x=\omega R_a$. The dotted line corresponds to $g_0=27/16$.}
	\label{fig:densityofstates}
\end{figure}

When considering $|t_{\omega l}|^2$, there are essentially two different regimes depending on whether the potential barrier created by the angular momentum $l$ is comparable or not to the frequency $\omega$ of the wave. 
In the high energy limit, $\omega \gg \kappa/(2\pi)$, one expects geometrical optics to be applicable thus yielding perfect absorption for $\omega \gtrsim l/R_a$ and perfect reflection for $\omega \lesssim l/R_a$, i.e. \cite{Page:1976df}
\begin{align}\label{thetawRl}
	\left| t_{\omega l} \right|^2 = \Theta(\omega R_a -l)
\end{align}
where $R_a=3\sqrt{3} r_s/2$ is the effective absorbing radius.
On the other hand, for low energies, $\omega  \ll \kappa/(2\pi)$, the probability of reflection will be higher for larger $l$. The expressions are more involved in this case but also known in closed form. For spinless particles they read \cite{Starobinsky:1973aij,Starobinskil:1974nkd,Page:1976df}
\begin{align}
	\left| t_{\omega l} \right|^2 =2 \left[\frac{l!^2}{(2l)!(2l+1)!}\right]^2 \left(\frac{\omega}{\kappa}\right) \left( 2 \omega \kappa r_s^2 \right)^{2l+1}  \prod_{n=1}^{l} \left[1+\left(\frac{\omega}{n\kappa-\kappa/2}\right)^2\right] \,,
\end{align}
so that the dominant contribution comes comes from the $l=0$ mode
\begin{align}
	\left| t_{\omega 0} \right|^2 =4 \omega^2 r_s^2  \,.
\end{align}
Therefore, the density of states in the two regimes only differs by
\begin{align}
	g_0 = \frac{g(\omega \gg \kappa/(2\pi)) }{g(\omega \ll \kappa/(2\pi))} \simeq  \frac{27}{16} \, .
\end{align}
Improved analytical expressions valid for intermediate high frequency have also been found in \cite{Sanchez:1976fcl,Sanchez:1976xm,Sanchez:1977vz}. Building on that work, a fitting formula for the density of states for the full range of frequencies has been obtained in \cite{Gray:2015xig} and it is given by
\begin{align}
	\frac{g(x)}{x^2} \simeq 1- \sqrt{2g_0}\, \text{sinc}(2\pi x )+ \left(\sqrt{2g_0}-\frac{11}{27} \right)  \text{sinc}\left(\frac{3\pi}{2}  x\right)  e^{-x^2} 
\end{align}
where $x=\omega R_a$ and $\text{sinc}(x)=\sin(x)/x$. This expression shows that the density of states does not vary much as a function of the frequency (c.f. Figure~\ref{fig:densityofstates}).

\bibliography{SuperHawking}
\bibliographystyle{JHEP}
\end{document}